\documentstyle[psfig,amstex]{mn}
 
\begin{document}

\title[The 2dF Galaxy Redshift Survey]{The 2dF Galaxy Redshift 
Survey: The Number and Luminosity Density of Galaxies}

\author[Cross et al.]{\parbox[t]{\textwidth}{Nicholas Cross$^1$, Simon P.\ Driver$^1$, Warrick 
Couch$^2$, Carlton M.\ Baugh$^3$, Joss Bland-Hawthorn$^4$, Terry Bridges$^4$, 
Russell Cannon$^4$, Shaun Cole$^3$, Matthew Colless$^5$, Chris Collins$^6$, 
Gavin Dalton$^7$, Kathryn Deeley$^2$, Roberto De Propris$^2$, 
George Efstathiou$^8$, Richard S.\ Ellis$^9$, Carlos S.\ Frenk$^3$, 
Karl Glazebrook$^{10}$, Carole Jackson$^5$, Ofer Lahav$^8$, Ian Lewis$^4$, 
Stuart Lumsden$^{11}$, Steve Maddox$^{12}$, Darren Madgwick$^8$, Stephen 
Moody$^8$, Peder Norberg$^3$, John A.\ Peacock$^{13}$, Bruce A.\ Peterson$^5$, 
Ian Price$^5$, Mark Seaborne$^7$, Will Sutherland$^13$, Helen Tadros$^7$, 
Keith Taylor$^4$}
\vspace*{6pt} \\ 
$^1$School of Physics and Astronomy, North Haugh, St Andrews, Fife, KY16 9SS,
United Kingdom \\
$^2$Department of Astrophysics, University of New South Wales, Sydney, NSW2052,
Australia \\
$^3$Department of Physics, South Road, Durham DH1 3LE, United Kingdom \\
$^4$Anglo-Australian Observatory, P.O. Box 296, Epping, NSW 2121, Australia \\
$^5$Research School of Astronomy \& Astrophysics, The Australian National 
University, Weston Creek, ACT 2611,Australia \\
$^6$Astrophysics Research Institute, Liverpool John Moores University, Twelve 
Quays House, Egerton Wharf, Birkenhead, L14 1LD, UK \\
$^7$Department of Physics, Keble Road, Oxford OX3RH, United Kingdom \\
$^8$Institute of Astronomy, University of Cambridge, Madingley Road, Cambridge 
CB3 0HA, United Kingdom \\
$^9$Department of Astronomy, 105-24, California Institute of Technology, 
Pasadena, CA, 91125, USA \\
$^{10}$Department of Physics and Astronomy, John Hopkins University, Baltimore,
MD, 21218, USA \\
$^{11}$Department of Physics \& Astronomy, E C Stoner Building, Leeds LS2 9JT,
United Kingdom \\
$^{12}$School of Physics and Astronomy, University of Nottingham, University 
Park, Nottingham, NG7 2RD, United Kingdom \\
$^{13}$Institute of Astronomy, University of Edinburgh, Royal Observatory, 
Edinburgh EH9 3HJ, United Kingdom \\
}

\maketitle

\begin{abstract}
We present the {\it bivariate brightness distribution} (BBD) for the 
2dF Galaxy Redshift Survey (2dFGRS) based on a preliminary subsample of 
45,000 galaxies. The BBD is an extension of the galaxy luminosity 
function incorporating surface brightness information. It allows the 
measurement of the local luminosity density, $j_{B}$, and the galaxy luminosity
and surface brightness distributions while accounting for surface brightness 
selection biases. 

The recovered 2dFGRS BBD shows a strong surface brightness-luminosity relation
($M_{B} \propto (2.4 \pm ^{1.5}_{0.5}) \mu_{e}$) providing a new constraint for
galaxy formation models. In terms of the number-density 
we find that the peak of the galaxy population lies at $M_B \geq -16.0$ mag. 
Within the well defined selection limits ($-24 < M_{B} < -16.0$ mag, 
$18.0 < \mu_e < 24.5$ mag arcsec$^{-2}$) the contribution towards the 
luminosity-density is dominated by conventional giant 
galaxies (i.e. 90\% of the luminosity-density is contained within 
$-22.5<M<-17.5$, $18.0<\mu_{e}<23.0$). The luminosity-density peak 
lies away from the selection boundaries implying that the 2dFGRS 
is complete in terms of sampling the local luminosity-density and that luminous
low surface brightness galaxies are rare.
The final value we derive for the local luminosity-density, inclusive of 
surface brightness corrections, is: 
$j_{B}=2.49 \pm 0.20 \times10^{8}h_{100}$L$_{\odot}$Mpc$^{-3}$.
Representative Schechter function parameters are: M$^*=-19.75\pm0.05$, 
$\phi^*=2.02\pm0.02\times10^{-2}$ and $\alpha=-1.09\pm0.03$.
Finally we note that extending the conventional methodology to 
incorporate surface brightness selection effects has resulted in an increase 
in the luminosity-density of $\sim 37$\%. Hence surface brightness selection
effects would appear to explain much of the discrepancy between previous 
estimates of the local luminosity density.

\end{abstract}

\section{Introduction}
Of paramount importance in determining the mechanism(s) and epoch(s) of
galaxy formation  (as well as the local luminosity density), is the accurate 
and detailed quantification of the local galaxy population. It represents
the benchmark against which both environmental and evolutionary 
effects can be measured. Traditionally this research area originated with the 
all-sky photographic surveys coupled with a few handfuls of hard earned 
redshifts. Over the past decade this has been augmented by both CCD-based 
imaging surveys and multi-slit/fibre-fed spectroscopic surveys. From these 
data, a number of perplexing problems have arisen, most notably: the faint 
blue galaxy problem (Koo \& Kron 1992; Ellis 1997), the local 
normalisation problem (Maddox {\it et al.} 1990; Shanks 1990; Driver, 
Windhorst \& Griffiths 1995; Marzke {\it et al.} 1998), the cosmological 
significance of low surface brightness galaxies (Disney 1976; McGaugh 1996;
Sprayberry {\it et al.} 1996; Dalcanton {\it et al.} 1997; Impey \& Bothun 
1997) and dwarf galaxies (Babul \& Rees 1992; Phillipps \& Driver 1995; 
Loveday 1997; Babul \& Ferguson 1996). These issues largely remain unresolved 
and arguably await an improved definition of the local galaxy population 
(Driver 1999). 

Recent advancements in technology now allow for wide field-of-view CCD 
imaging surveys\footnote{It is sobering to note that the largest published 
CCD-based imaging survey to date is 17.5 sq degrees to a central surface 
brightness of 25 V mag arcsec$^{-2}$ (Dalcanton {\it et al.} 
1997) as compared to the all sky 
coverage of photographic media} and bulk redshift surveys through purpose 
built multi-fibre spectrographs such as the common-user two-degree field (2dF) 
facility at the Anglo Australian Telescope (Taylor, Cannon \& Parker 1998). 
The Sloan Digital Sky Survey elegantly combines these two facets (Margon 
1999).

The quantity and quality of data that is becoming available allows not only 
the revision of earlier results but more fundamentally the opportunity to 
review and enhance the methodology with which the local galaxy population is
represented. For instance some criticism that might be levied at the
current methodology --- the representation of the space density of 
galaxies using the Schechter Luminosity function (Schechter 1976; Felten 1985;
Binggeli, Sandage \& Tammann 1988) --- is that 
firstly, it assumes that galaxies are single parameter systems defined
by their apparent magnitude alone, and secondly it describes the entire galaxy 
population by only three parameters; the characteristic luminosity $L_{*}$, 
the normalisation of the characteristic luminosity $\phi_{*}$, and the 
faint-end slope $\alpha$. While it is desirable to represent the population 
with the minimum number of parameters, important information may lie in the 
nuances of detail.

In particular, two recent areas of research suggest a greater diversity in the 
galaxy population than is allowed by the Schechter function form. Firstly, 
Marzke {\it et al.} (1994) and also Loveday (1997) report the indication of a 
change in the faint end slope at faint absolute magnitudes --- a possible 
giant-dwarf transition --- and this is also seen in a number of Abell clusters 
where it is easier to probe into the dwarf regime (e.g., Driver {\it et al.}
1994; De Propris {\it et al.} 1995; Driver, Couch \& Phillipps 1998; Trentham
1998). Secondly a number of studies show that the three Schechter parameters, 
and in particular the faint-end slope, 
have a strong dependence upon: surface brightness limits 
(Sprayberry, Impey \& Irwin 1996; Dalcanton 1998); colour (Lilly 
{\it et al.} 1996); spectral type (Folkes {\it et al.} 1999); optical 
morphology (Marzke {\it et al.} 1998), environment (Phillipps {\it et al.}
1998) and wavelength (Loveday 2000). It has been noted (Willmer 1997) that 
the choice of method for reconstructing the galaxy LF also contains some 
degree of bias.

More fundamentally, evidence that the current methodology might actually be 
flawed comes from comparing
recent measurements of the galaxy luminosity function as shown in Fig. 1.
The discrepancy between these surveys is significantly adrift from the quoted 
formal errors implying an unknown systematic error. The range of discrepancy 
can be quantified as a factor of 2 at the $L_{*}$ ($M \sim -19.5$) point 
rising to a factor of 10 at $0.01L_{*}$ ($M \sim -14.5$). The impact of 
this variation is a factor of 3-4, for instance, in assessing the 
contribution of galaxies to the local baryon budget 
(e.g. Persic \& Salucci 1992; Bristow \& Phillipps 1994; and Fukugita, Hogan \&
Peebles 1998). 

\begin{figure}
\vspace{-1.0cm}
{\psfig{file=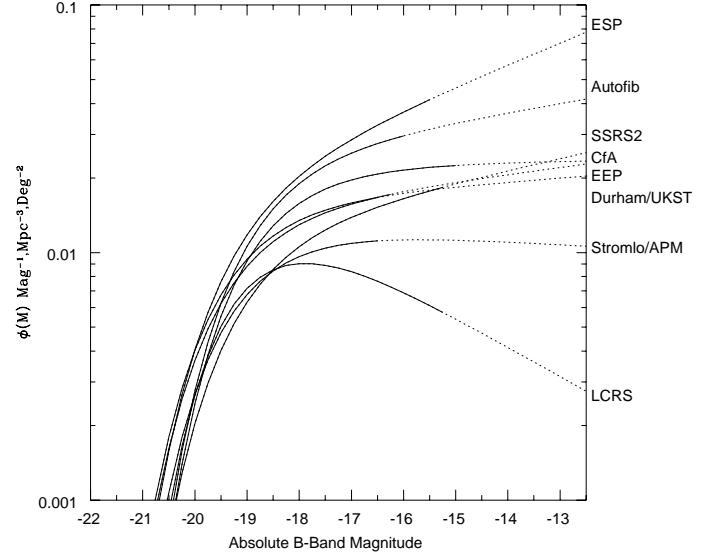,width=90mm,height=90mm}}
\caption{Schechter luminosity functions from recent magnitude-limited
redshift surveys (q.v. Table 1). The line becomes dotted outside the range of 
survey data. The range of values show the uncertainty in the LF which in turn 
filters through to the local measure of the mean luminosity density.}
\end{figure}

This uncertainty is in addition to that introduced from the unanswered 
question of the space density of low surface brightness galaxies. The
most recent attempt to quantify this is by O'Neil \& Bothun (2000) ---
following on from McGaugh (1996), and in turn Disney (1976) --- 
who conclude that the surface brightness function (SBF) of galaxies --- 
the number density of galaxies in intervals of surface brightness --- 
is of similar form to 
the luminosity function. Thus both the LF and SBF are described by a flat 
distribution with a cutoff at bright absolute magnitudes or high surface 
brightnesses. Taking the O'Neil result at face value, this implies 
a further error in measures of the local luminosity density of 2-3 - i.e. the
contribution to the luminosity- (and hence baryon-) 
density from galaxies is uncertain to a factor
of $\sim 10$. However the significance of low surface brightness
galaxies depends upon their luminosity range and similarly the 
completeness of the LF relies on the surface brightness intervals over which 
each luminosity bin is valid. Both representations are incomplete unless
the information is combined. This leads 
us to the conclusion that both the total
flux and the manner in which this flux is distributed must be dealt with
simultaneously. Several papers have been published which deal with either
surface brightness distributions or Bivariate Brightness Distributions
(Phillipps \& Disney 1986, Boyce \& Phillipps 1994, Minchin 1999, Sodre \& 
Lahav 1993, and Petrosian 1998). These are either theoretical, limited to 
cluster environments or have poor statistics due to the scarcity of good 
redshift data.

Recently, Driver (1999) determined the first measure of the
{\it bivariate brightness distribution} for field galaxies using Hubble Deep 
Field data (Williams {\it et al.} 1996) and capitalising on photometric 
redshifts (Fern\'andez-Soto {\it et al.} 1998). The result, based on a
volume limited sample of 47 galaxies, implied that giant low surface brightness
galaxies were rare but that there exists a strong Luminosity-Surface Brightness
relationship, similar to that seen in Virgo (Binggeli 1993). The sense of the
relationship  implied that low surface brightness galaxies are preferentially 
of lower luminosity (i.e. dwarfs). If this is borne out it strongly tempers the
conclusions of O'Neil \& Bothun (2000). While the number of low surface 
brightness galaxies may be large, their luminosities are low, so
their contribution to the local luminosity density, is also low, $< 20$\% 
(Driver 1999).

This paper attempts to bundle these complex issues onto a more intuitive
platform by expanding the current representation of the local galaxy population
to allow for: surface brightness detection effects, star-galaxy separation
issues, surface brightness photometric corrections and clustering effects. 
This is achieved by expanding the mono-variate luminosity function
into a bivariate brightness distribution (BBD) where the additional 
dimension is surface brightness. The 2dFGRS allows us to do this for the
first time by having a large enough database to separate galaxies in both
magnitude and surface brightness without having too many problems with small
number statistics.

In \S 2 we discuss the revised methodology for measuring the space density
of the local galaxy population, the local luminosity density and the 
contribution towards the baryon density in detail. In \S 3 we present
the current 2dFGRS data (containing $\sim 45,000$ galaxies or one fifth of 
the expected final tally). In \S 4 we 
correct for the light lost under the isophote and define our surface
brightness measure. In \S 5 we apply the methodology to construct the 
first statistically significant bivariate brightness distribution for field 
galaxies. The results for the number-density and luminosity-density are 
detailed in \S 6 and \S 7. In \S 8, we compare these results to other 
surveys. Finally we present our conclusions.

Throughout we adopt $H_{o} = 100$kms$^{-1}$Mpc$^{-1}$ and a standard flat
cosmology with zero cosmological constant ({\it i.e.} $q_{o}=0.5, \Lambda=0$).
However we note that the results presented here are only weakly dependent on 
the cosmology.

\section{Methodology}
The luminosity density, $j$, is the total amount of flux 
emitted by all galaxies per Mpc$^{3}$. When measured in the UV band it
can be converted to a measure of the star-formation rate (see for example 
Lilly {\it et al.} 1996, Madau {\it et al.} 1998). When measured at longer 
wavelengths it can be combined with mass-to-light ratios to yield an 
{\it approximate} value for the contribution from galaxies towards the local 
matter density $\Omega_{M}$ --- independent of $H_{o}$, only weakly cosmology
dependent and not reliant on any specific theory of structure formation 
(see for example Carlberg {\it et al.} 1996; Fukugita, Hogan \& Peebles 1998).
The two main caveats are firstly the accuracy of $j_{B}$ (the luminosity 
density measured in the B-band), and secondly the 
assumption of a ubiquitous mass-to-light ratio. 

\subsection{Measuring $j$}

The luminosity density, $j$, is found by integrating the product of the 
number density $\Phi(L/L_{*})$ and the luminosity L with respect to luminosity.
\begin{equation}
j=\int^{\infty}_{0}L\,\Phi(L/L_{*})\,d(L/L_{*}) \label{eq:j}
\end{equation}
By convention, $j$ is typically derived
from a magnitude-limited redshift survey by determining the representative
Schechter parameters for a survey (e.g. Efstathiou {\it et al.} 1988) and 
then integrating the luminosity weighted Schechter function, 
where $\Phi(L/L_{*})\,d(L/L_{*})$ is the Schechter function 
(Schechter 1976) given by:
\begin{equation}
\Phi(L/L_{*})\,d(L/L_{*})=\phi_{*}\,(L/L_{*})^{\alpha}\,\exp{[-(L/L_{*})]}\,d(L/L_{*})
\end{equation}
and $\phi_*, L_{*},$ and $\alpha$ are the three parameters which define 
the survey (referred to as the normalisation point, characteristic turn-over 
luminosity and faint-end slope parameter respectively).
More simply if a survey is defined by these three parameters it follows
that:
\begin{equation}
j=\phi_{*}\,L_{*}\,\Gamma(\alpha+2)
\end{equation}

Table 1 shows values for the luminosity density derived
from a number of recent magnitude-limited redshift surveys (as indicated).
The variation between the measurements of $j$ from these surveys is 
$\sim 2$ and hence the uncertainty in the galaxy contribution to the mass 
budget is at best equally uncertain. This could be due to a number of 
factors, e.g., large scale-structure, selection biases,
redshift errors, photometric errors or other incompleteness. In this paper 
we wish to explore the possibility of selection bias due to surface brightness
considerations only. The principal motivation for this is that the LCRS
(lower line on Fig. 1), which recovers the lowest $j$ value, adopted a 
bright isophotal detection limit of $\mu_{r} = 23$ mag arcsec$^{-2}$, 
suggesting a dependence between the measured $j$
and the surface brightness limit of the survey. Here we develop a method
for calculating $j$ which incorporates a number of 
corrections/considerations for surface brightness selection biases.
In particular, a surface brightness dependent Malmquist correction, a
surface brightness redshift completeness correction and an isophotal 
magnitude correction. We also correct for clustering. What is not included 
here, and will be pursued in a later paper, is the photometric accuracy, 
star-galaxy separation accuracy and a detection correction specifically for 
the two degree field galaxy redshift survey (2dFGRS).

Implementing these corrections requires re-formalising the path to 
$j$. Firstly, we replace the luminosity function representation
of the local galaxy population by a {\it bivariate brightness distribution}
(BBD). The bivariate brightness distribution is the 
galaxy number density, $\Phi$, as a function of absolute, total, B-band 
magnitude, $M_{B}$, and absolute, effective surface brightness, 
$\mu_e$, {\it i.e.}, $\Phi(M,\mu)$. 
To construct a BBD we need to convert the observed distribution to a
number density distribution taking into account the Malmquist bias
and the redshift incompleteness correction, {\it i.e.}, 
\begin{equation}
\Phi(M,\mu)=\frac{O(M,\mu)+I(M,\mu)}{V(M,\mu)}\,W(M,\mu) 
\label{eq:phiM}
\end{equation}
where:

\begin{itemize}
\item $O(M,\mu)$ is the matrix of absolute magnitude, $M$, and absolute 
effective surface brightness, $\mu$, for galaxies with redshifts. 

\item $I(M,\mu)$ is the matrix of absolute magnitude, $M$, and absolute 
effective surface brightness, $\mu$, for those galaxies for which redshifts 
were not obtained.

\item $V(M,\mu)$ is the matrix which specifies the volume over which a 
galaxy with absolute magnitude, $M$, and absolute effective surface 
brightness, $\mu$, can be seen (see also Phillipps, Davies \& Disney 1990). 

\item $W(M,\mu)$ is the matrix that weights each bin to compensate
for clustering.

\end{itemize}
Deriving these matrices is discussed in detail later.
$j$ is then defined as:
\begin{equation}
j=\sum^{M} \sum^{\mu} L(M)\,\Phi(M,\mu)
\end{equation}
or in practice,
\begin{equation}
j=\sum^{M} \sum^{\mu} 
10^{-0.4(M-M_{\odot})}\,\Phi(M,\mu)
\end{equation}
in units of $L_{\odot}$Mpc$^{-3}$ where $M_{\odot\,\rm B} = 5.48$.

~

Our formalism has two key advantages over the traditional luminosity 
function: Firstly, it adds the additional dimensionality of surface 
brightness allowing for surface brightness specific corrections. Secondly, 
it represents the galaxy population by a distribution rather than a function,
thus requiring no fitting procedures or assumption of any underlying parametric
form. 

\section{The Data}
The data set presented here is based upon a sub-sample of the Automated Plate 
Measuring-machine galaxy catalogue (APM; Maddox et al. 1990a,b) for which
spectra have been obtained using the two-degree field Facility (2dF) at the 
Anglo-Australian Telescope (AAT). 

The original APM catalogue contains $b_J$ magnitudes with random error
$\Delta m = \pm 0.2$ mag (Folkes et al. 1999) and isophotal areas 
$A_{\rm ISO}$. The isophotal 
area is defined as the number of pixels above a limiting isophote, 
$\mu_{\rm lim}$ --- set at the 
$2 \sigma$ level above the sky background ($\mu_{lim} \approx 
24.53$ mag arcsec$^{-2}$ with a variation of $\pm0.11$ mag arcsec$^{-2}$ - 
Pimbblet et al. 2000). However APM $b_J$ magnitudes are found to vary from CCD
$b_J$ magnitudes by $0.14\pm0.29$mag (Metcalfe et al. 1995). Therefore the 
isophotal limit in APM $b_J$ magnitudes is $\mu_{lim} = 24.67\pm0.30$mag 
arcsec$^{-2}$. 
One pixel equals 0.25 $\Box''$. The minimum isophotal area 
found for galaxies in the APM catalogue is 35$\Box''$. Star-galaxy separation
was implemented as described in Maddox {\it et al.} (1990b). The final APM 
sample contains $3\times10^{6}$ galaxies covering 15,000 square degrees -- 
see Maddox {\it et al.} (1990a,b) for further details.

The 2dFGRS input catalogue is a 2000 $\Box ^o$ subregion of the APM catalogue
(covering two continuous regions in the northern and southern Galactic caps
plus random fields)
with an extinction corrected magnitude limit of $m=19.45$ (Colless 1999).
The input catalogue contains 250,000 galaxies for which 81,895 have been 
observed
using 2dF (as of November 1999). Each spectrum within the database has 
been examined by eye to check if the redshift is reliable. Redshifts 
are determined via cross-correlation with specified templates (see Folkes 
{\it et al.} 1999 for details). 
A brief test of the reliability of the 2dFGRS was achieved via a comparison
between 1404 galaxies in common with the Las Campanas Redshift Survey 
(Lin {\it et al.} 1996) for which there were only 8 mismatches, showing that 
2dF redshifts are reliable. Of the 81,895 galaxies, 74,562 have a redshift, 
resulting in a redshift completeness of 91\%.

The survey comprises many overlapping two degree fields and many still have to
be observed. Hence the absolute normalisation is tied to the full input galaxy
catalogue which is known to contain 174.0 galaxies with $m \leq 19.45$ 
per square degree. Using a subsample of 44796 galaxies, covering just the
South Galactic Pole (SGP) region, this yields an {\it effective} coverage for 
this survey of 257 non-contiguous square degrees. 

Finally, for the purposes of this paper, we adopt a lower redshift limit of
$z = 0.015$ to minimise the influence of peculiar velocities in the
determination of absolute parameters and an upper redshift limit of $z = 0.12$.

This upper limit of $z=0.12$ was selected so as to maximise the sample size 
yet minimise the error introduced by the isophotal corrections. At $z=0.12$ 
the uncertainty in the isophotal correction ($\pm^{0.07}_{0.16}$), due to
type uncertainty (see Appendix A), remains smaller than the photometric error
($\pm 0.2$ mags). Note that the increase in the error in the isophotal 
correction is primarily 
because of the increase in the {\it intrinsic} isophotal limit 
due to a combination of surface brightness-dimming and the K-correction. 

The final sample is therefore pseudo {\it volume}-limited and 
contains 20,765 galaxies, with redshifts, selected from a parent sample of 
45,000.

~

\noindent
Note: all magnitude and surface brightnesses are in the APM $b_{j}$ filter.

\section{Isophotal corrections}

The APM magnitudes have already been corrected assuming a Gaussian 
profile (see Maddox {\it et al.} 1990b for full details). This was 
aimed primarily at recovering the light lost due to the seeing and
is crucial for compact objects. It is known to significantly 
underestimate the isophotal correction required for low surface 
brightness disks. Such systems typically exhibit exponential profiles with
disks which can extend a substantial distance beyond the isophote,
the most famous example being Malin 1 (Bothun {\it et al.} 1987). Once
thought of as a Virgo dwarf this system remains the most luminous 
field galaxy known. 

To compliment the Gaussian correction (required for compact objects
but ineffectual for extended sources) we introduce an additional
correction (ineffectual for compact sources but suitable for extended
disks). This correction assumes all objects can be represented by
a pure exponential surface brightness profile extending from the
core outwards. In this case the surface brightness profile is simply:
\begin{equation}
\Sigma(r) = \Sigma_o \exp(-r/\alpha)
\end{equation}
or,
\begin{equation}
\mu(r) = \mu_o + 1.086(r/\alpha)  \label{eq:exp}
\end{equation}
Where $\Sigma_o$ is the central surface brightness in Wm$^{-2}$ arcsec$^{-2}$, 
$\alpha$ is the scale length of the galaxy in arcsecs and r the radius in 
arcsecs. $\mu_o$ is the central surface brightness in mag arcsec$^{-2}$.

Under this assumption a galaxy's observed isophotal luminosity is
the integrated radial profile out to $r_{iso}$.
\begin{equation}
l_{iso}=2 \pi \int^{r_{iso}}_0 \Sigma_o \exp(-r/\alpha) rdr
\end{equation}
which can be expressed in magnitudes as:
\begin{equation}
m_{iso}=\mu_o^{app}-2.5\log_{10} \{2\,\pi[\alpha^2 
-\alpha(\alpha+r_{iso})\exp(-r_{iso}/\alpha)]\} \label{eq:miso}
\end{equation}
(here $\mu^{app}_{o}$ denotes the apparent surface brightness
uncorrected for redshift.)
$\mu_{lim}$, the detection/photometry isophote, can be expressed as:
\begin{equation}
\mu_{lim}=\mu_o^{app} + 1.086(r_{iso}/\alpha) \label{eq:mulim}
\end{equation}
As $m_{iso}$, $r_{iso}$ and $\mu_{lim}$ are directly 
measurable quantities, equations (~\ref{eq:miso} \& ~\ref{eq:mulim} ) can be 
solved numerically. The total magnitude is then given by:
\begin{equation}
l_{tot}=2 \pi \int^{\infty}_0 \Sigma_o \exp(-r/\alpha) rdr = 2 \pi \Sigma_{o} \alpha^{2}
\end{equation}
or,
\begin{equation}
m^{tot}=\mu_o^{app}-2.5\log_{10} (2 \pi \alpha^2)
\end{equation}

From this description an extrapolated central surface
brightness can be deduced numerically from the specified isophotal area and 
isophotal magnitude (after the seeing correction).
Note that this prescription ignores the possible presence of a bulge, opacity,
and inclination leading to an underestimate of the isophotal correction. 
This is unavoidable as the data quality is insufficient to establish 
bulge-to-disk ratios. To verify the impact of this we explore the accuracy
of the isophotal correction for a variety of galaxy types in Appendix A.
The tests show that the isophotal correction 
is a significant improvement over the isophotal magnitudes for all types - 
apart for ellipticals where the introduced error is negligible compared to the
photometric error - and a dramatic improvement for low surface brightness 
systems. The final magnitudes, after isophotal correction, now lie well within 
the quoted error of $\pm 0.2$ mags for both high- and low-surface brightness
galaxies.

\subsection{The Effective Surface Brightness}

Most results cited in the literature use the central surface brightness or 
the effective surface brightness. The central surface brightness, as
described above, is the extrapolated surface brightness at the core under
the assumption of a perfect exponential disk. The effective surface brightness
is the mean surface brightness within the half-light radius. The conversion 
between the measures is relatively straightforward and described as follows:
\begin{equation}
l_{\frac{1}{2}}=\pi\,\Sigma_o\,\alpha^2=2\pi\,\Sigma_o\,\alpha^2[1-(1+r_e/\alpha)\exp(-r_e/\alpha)]
\end{equation}
which can be solved numerically to get
\begin{equation}
r_e=1.678 \alpha
\end{equation}
The effective surface brightness is now given by:
\begin{equation}
\mu_e^{app}=\mu_o^{app}+2.5\log_{10}[(r_e/\alpha)^2]=\mu_o^{app}+1.124
\end{equation}
Hence from the isophotal magnitudes and areas we can
derive the total magnitude and effective
surface brightness (quantities which are now independent of the
isophotal detection threshold). We chose to work with effective surface 
brightness as it can, at some later stage, be measured directly from higher 
quality CCD data. 
Note that these surface brightness measures are all
{\it apparent} rather than {\it intrinsic}, however this is not important as
although surface brightness is distance dependent the isophotal correction 
is not (this is because both $\mu_{lim}$ and $\mu_{e}$ vary with 
redshift in the same way).

Fig. 2 shows the final 2dFGRS sample ({\it i.e.} after isophotal correction)
for those galaxies with (upper panel) and without (lower panel) redshifts. 
The galaxies are plotted according to their apparent total magnitude and 
apparent effective surface brightness. The curved boundary at the faint 
end of both plots is due to the 
isophotal corrections which are strongly dependent on $\mu_e$ for a constant 
$m$. As $\mu_e-1.124$ 
tends towards $\mu_{lim}$, the isophotal limit, the correction 
tends towards infinity, making it impossible to see galaxies close to 
$\mu_{lim}$. The average isophotal correction is $0.33$ mag (for 
$\mu_{lim} = 24.67$ mag arcsec$^{-2}$).

\begin{figure}
\centerline{\psfig{file=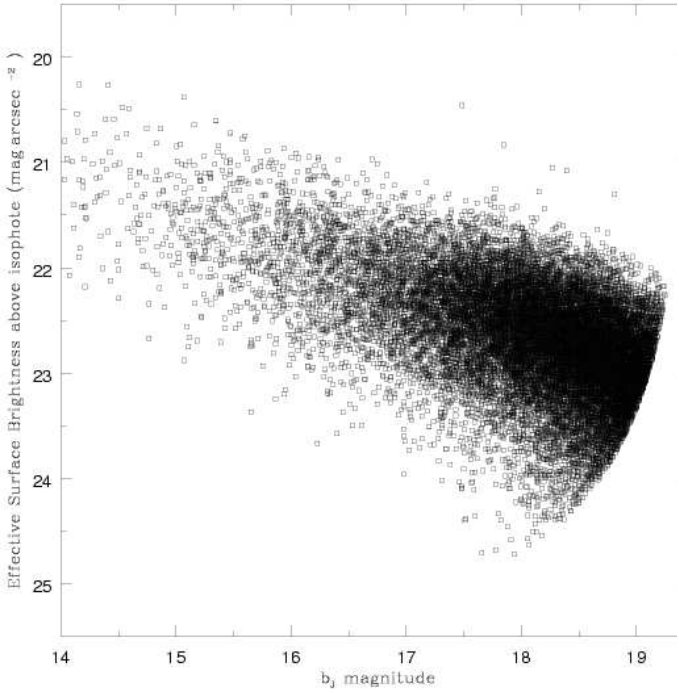,height=100.0mm,width=100.0mm}}
\centerline{\psfig{file=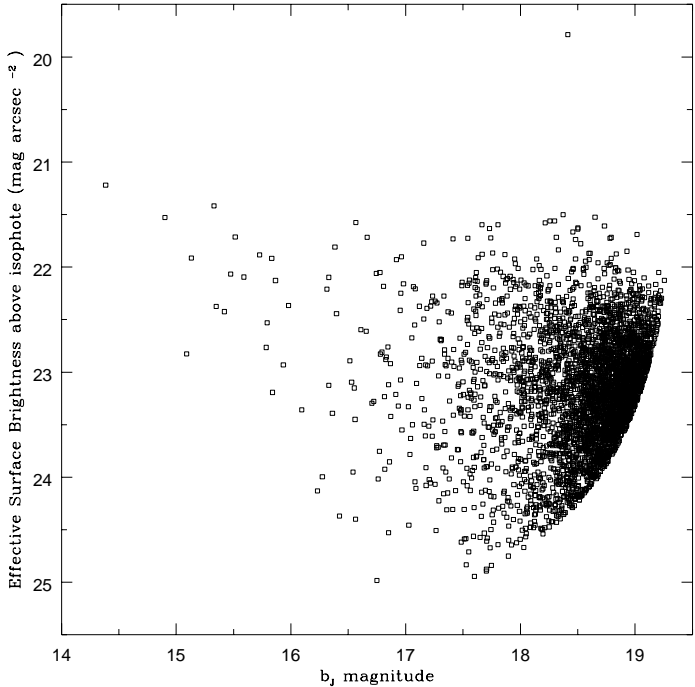,height=100.0mm,width=100.0mm}}
\caption{Galaxies with (upper) and without (lower) redshifts for the current
2dFGRS sample plotted according to their apparent extinction corrected total
magnitude and apparent effective surface brightness.}
\end{figure}

The observed mean magnitude and observed mean effective surface brightness for 
those galaxies with and without redshifts are: $18.06\pm0.01$ mag 
\& $22.66\pm0.01$ mag arcsec$^{-2}$ and $18.54\pm0.01$ mag \& $23.17\pm0.01$ 
mag arcsec$^{-2}$, respectively. These figures imply that galaxies closer to 
the detection limits are preferentially under-sampled. 

\section{Constructing the BBD}
We now apply the methodology described in section 2 to derive the BBD from our
data set. This requires constructing the four matrices, $O(M,\mu)$,
$I(M,\mu)$, $V(M,\mu)$ and $W(M,\mu)$.

\subsection{Deriving $O(M,\mu)$}
For those galaxies with redshifts, we can obtain their absolute
magnitude and absolute effective
surface brightness assuming a cosmological framework 
and a global K-correction\footnote{Individual K-corrections will
be derived from the data; however this has not yet been implemented.},
K(z)=2.5z (Driver {\it et al.} 1994). The conversions from observed 
to absolute parameters are given by:
\begin{equation}
M=m-5\log_{10}[\frac{2c}{H_{o}}(1+z)(1-(1+z)^{-0.5})]-25-K(z) \label{eq:absmag}
\end{equation}
and,
\begin{equation}
\mu_e=\mu_e^{app}-10\log_{10}(1+z)-K(z) \label{eq:absmu}
\end{equation}

Here $H_o$ is the Hubble constant, $c$ is the speed of light, $\mu_e^{app}$ is
the apparent effective surface brightness and $\mu_e$ is the absolute 
effective surface brightness. The $M$, derived by Eqn~\ref{eq:absmag} is 
a {\it total} absolute magnitude since the correction has been made for the 
light below the isophote. 

\begin{figure}
\centerline{\psfig{file=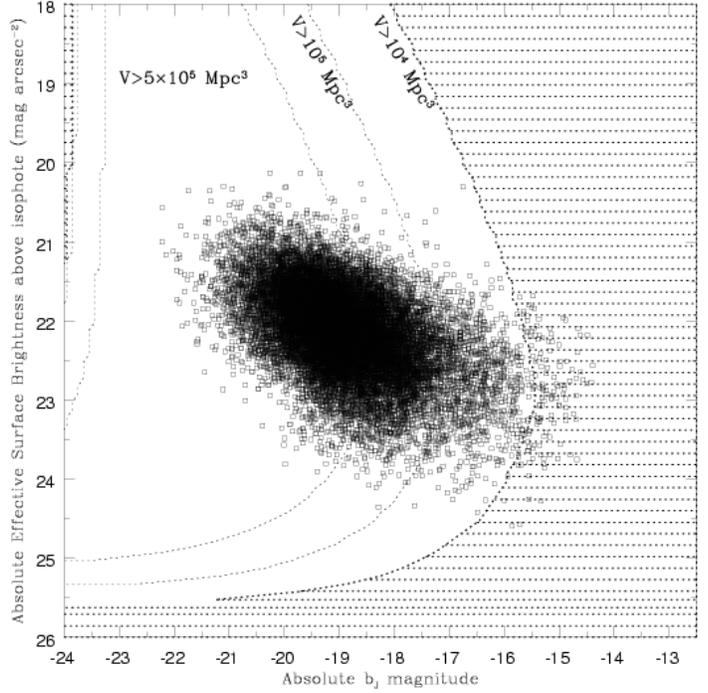,height=100.0mm,width=100.0mm}}
\caption{Galaxies from the 2dFGRS with redshifts, plotted in absolute
magnitude and absolute effective surface brightness space. The shaded region 
denotes the regions where less than $10^4$ Mpc$^{3}$ are surveyed and is 
based on visibility theory as described in Phillipps, Davies \& Disney (1990).
The 3 curves represent the volumes of $10^4$, $10^5$ and $5\times10^5$ 
Mpc$^{3}$.}
\end{figure}

Fig. 3 shows the upper panel of Fig. 2 with the axes converted to absolute 
parameter space using the conversions shown above. Naturally galaxies in
different regions are seen over differing volumes, because of Malmquist bias,
hence it is not yet valid
to compare the relative numbers. However it is possible to define lines
of constant volume as shown on Fig. 3 (dotted lines). These lines are derived
from visibility theory (Phillipps, Davies \& Disney 1990) and they delineate
the region of the BBD plane where galaxies can be seen over various volumes. 
The shaded region shows the region where the volume is less than 10$^{4}$ 
Mpc$^{3}$ and hence where we are insensitive to galaxy densities of $<  
10^{-2}$galaxies/Mpc$^{-3}$mag$^{-1}$(mag arcsec$^{-2}$)$^{-1}$. 
The equations used to calculate the lines are laid out in Appendix B. We 
show a $V=5\times10^5$Mpc$^{3}$ line rather than $V=10^6$Mpc$^{3}$ because the
$z=0.12$ limit is at a volume less than $V=10^6$Mpc$^{3}$.
The parameters used in the visibility calculations are:
$\mu_{lim}=24.67$ mag arcsec$^{-2}$; 
$\theta_{min}=7.2''$, $\theta_{max}=200.0''$, $m_{bright}=14.00$ mag; 
$m_{faint}=19.45$ mag; $z_{min}=0.015$; and $z_{max}=0.12$.
The clear space between the data and the selection boundary at bright 
absolute magnitudes implies that although the 2dFGRS data set samples a 
sufficiently representative volume ($V > 10,000$ Mpc$^{3}$), galaxies only 
exist over a restricted region of this observed BBD.
Fainter than $M = -16.5$ mags the volume is insufficient
to sample populations with a space density of 
$10^{-2}$ Mpc$^{-3}$mag$^{-1}$(mag arcsec$^{-2}$)$^{-1}$ or less.

Fig. 4 shows the data of Fig. 3 binned into $\mu_e$ and $M$ intervals to 
produce the matrix $O(M,\mu)$ (see Eqn~\ref{eq:phiM}). The bins are $0.5$ mag$ 
\times 0.5$ mag arcsec$^{-2}$ and start from $-24.0$mag and a central surface 
brightness of $19.0$mag arcsec$^{-2}$, effective surface brightness of 
$20.12$mag arcsec$^{-2}$. The total number of bins is 200 in a uniform  
$20\times10$ array. Fig. 4 represents the {\it observed} distribution of 
galaxies and shows a strong peak close to the typical $M_{*}$ value seen in 
earlier surveys (see references listed in Table 1).  

\begin{figure}
\centerline{\psfig{file=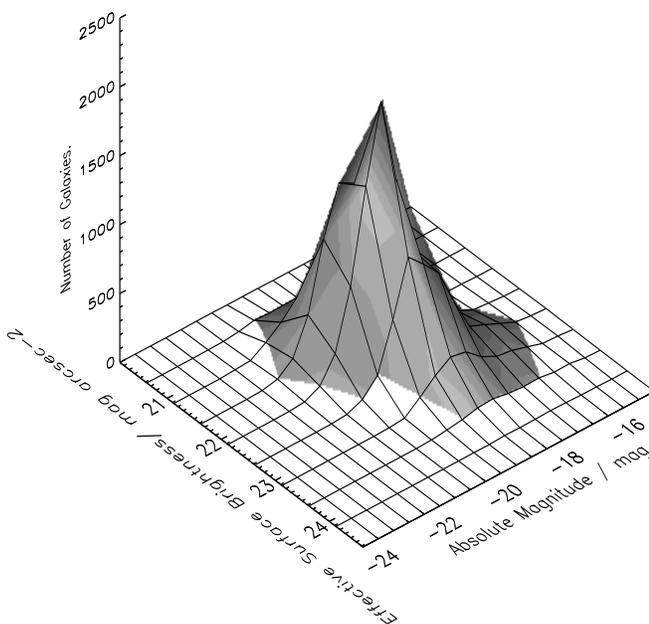,height=100.0mm,width=85.0mm}}
\centerline{\psfig{file=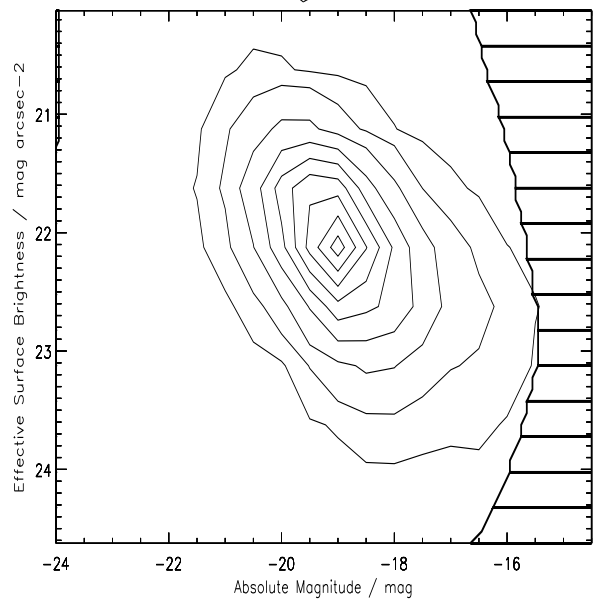,height=85.0mm,width=85.0mm}}
\caption{The observed distribution of the 2dFGRS data set mapped onto the
BBD, prior to volume and incompleteness correction. The contour lines are set 
at 25, 100, 250, 500, 750, 1000, 1250, 1500 and 1750 galaxies per bin. The 
minimum number of galaxies in a bin is 25.}
\end{figure}

\subsection{Deriving $I(M,\mu)$}
Not all galaxies targeted by the 2dFGRS have a measured redshift.
This may be due to lack of spectral features, selection biases or a 
misplaced/defunct fibre. One method to correct for these missing galaxies is 
to assume that they have the same observed
BBD as those galaxies for which redshifts
have been obtained. One can then simply scale up all bins by this known
incompleteness. 

However, from Fig. 2 and \S 1 \& 3 we noted 
that the incompleteness is a function
of both the apparent magnitude and the apparent surface brightness.
There is no reliable way of converting these values to absolute values without
redshifts and to obtain an incompleteness correction, $I(M,\mu)$, some 
assumption must be made. Here we assume that a galaxy of unknown redshift with 
apparent magnitude, $m$, and apparent effective surface brightness, 
$\mu_e$, has a range of possible BBD bins that can be {\it statistically} 
represented by the BBD distribution of galaxies 
with $m \pm \Delta m$ and $\mu_e \pm \Delta \mu_e$. 
The underlying assumption is that galaxies with and 
without redshifts with similar observed $m$ and $\mu$ have similar redshift 
distributions. I.e., the detectability of a galaxy is primarily dependent on 
its apparent magnitude and apparent surface brightness. [While these factors 
are obviously crucial one could also argue that additional factors, not 
incorporated here, such as the predominance of spectral features are also
important such that the true probability distribution for the missing galaxies 
could be somewhat skewed from that derived in Fig. 5.]

Hence for each galaxy, 
without a redshift, we select those galaxies, with redshifts, within $0.1$mag 
and $0.1$mag arcsec$^{-2}$ and determine their collective observed BBD 
distribution. This is achieved using all 44796 galaxies in the SGP sample, as 
galaxies, without redshifts, are not limited to $z<0.12$. This distribution is
then normalised to unity to generate a probability distribution for this 
galaxy. This is repeated for every galaxy without a redshift.
Fig. 5, shows the probability distribution for a galaxy 
with $m=17.839$ and $\mu_e=22.314$. There are 317 galaxies 
with redshifts within $0.1$mag and $0.1$mag arcsec$^{-2}$ and combined they 
have a distribution ranging from $M = -15.44$ to $M = -21.76$
and $\mu_e = 22.26$ to $\mu_e = 21.06$. 

\begin{figure}
\centerline{\psfig{file=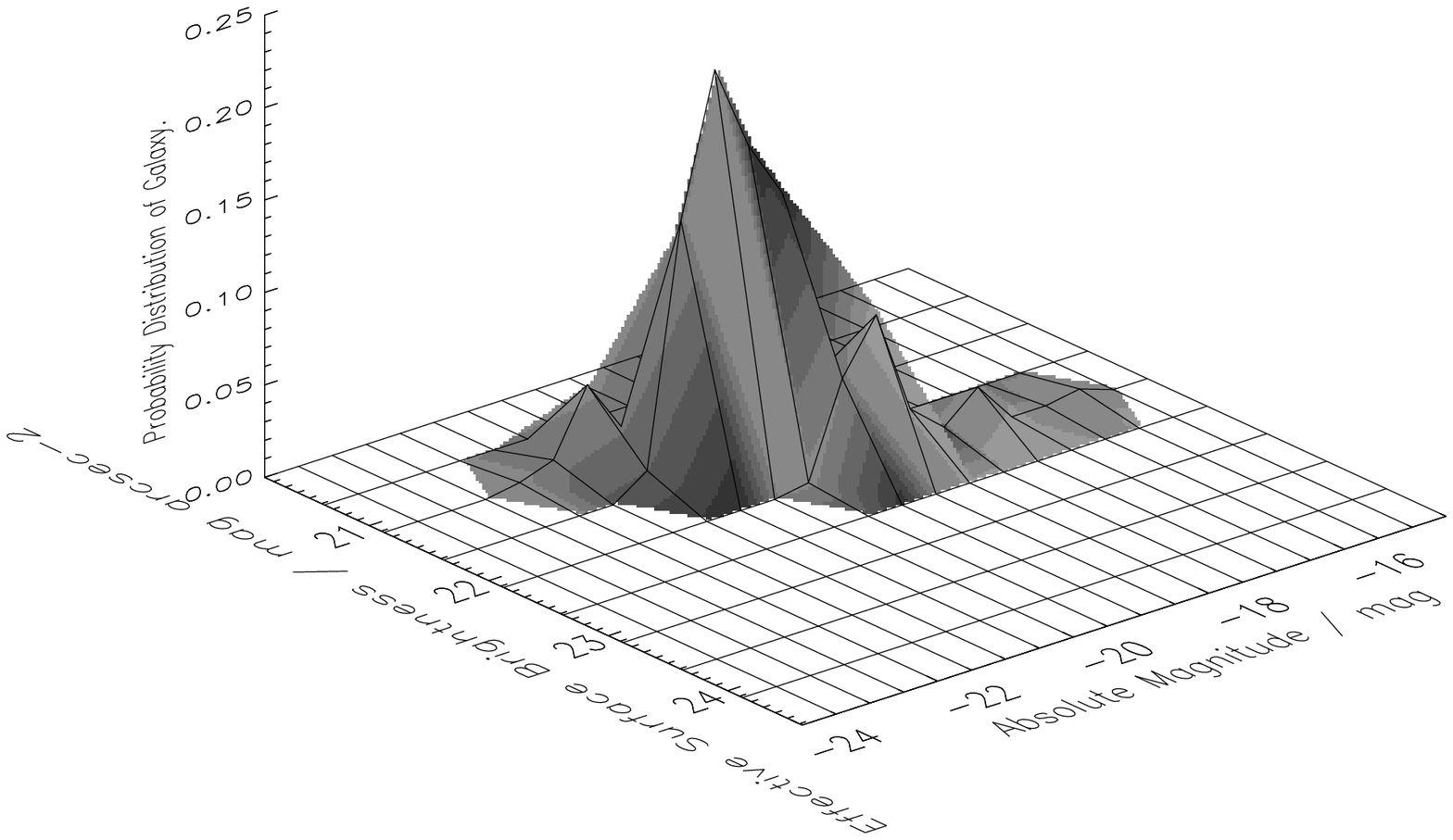,height=100.0mm,width=85.0mm}}
\centerline{\psfig{file=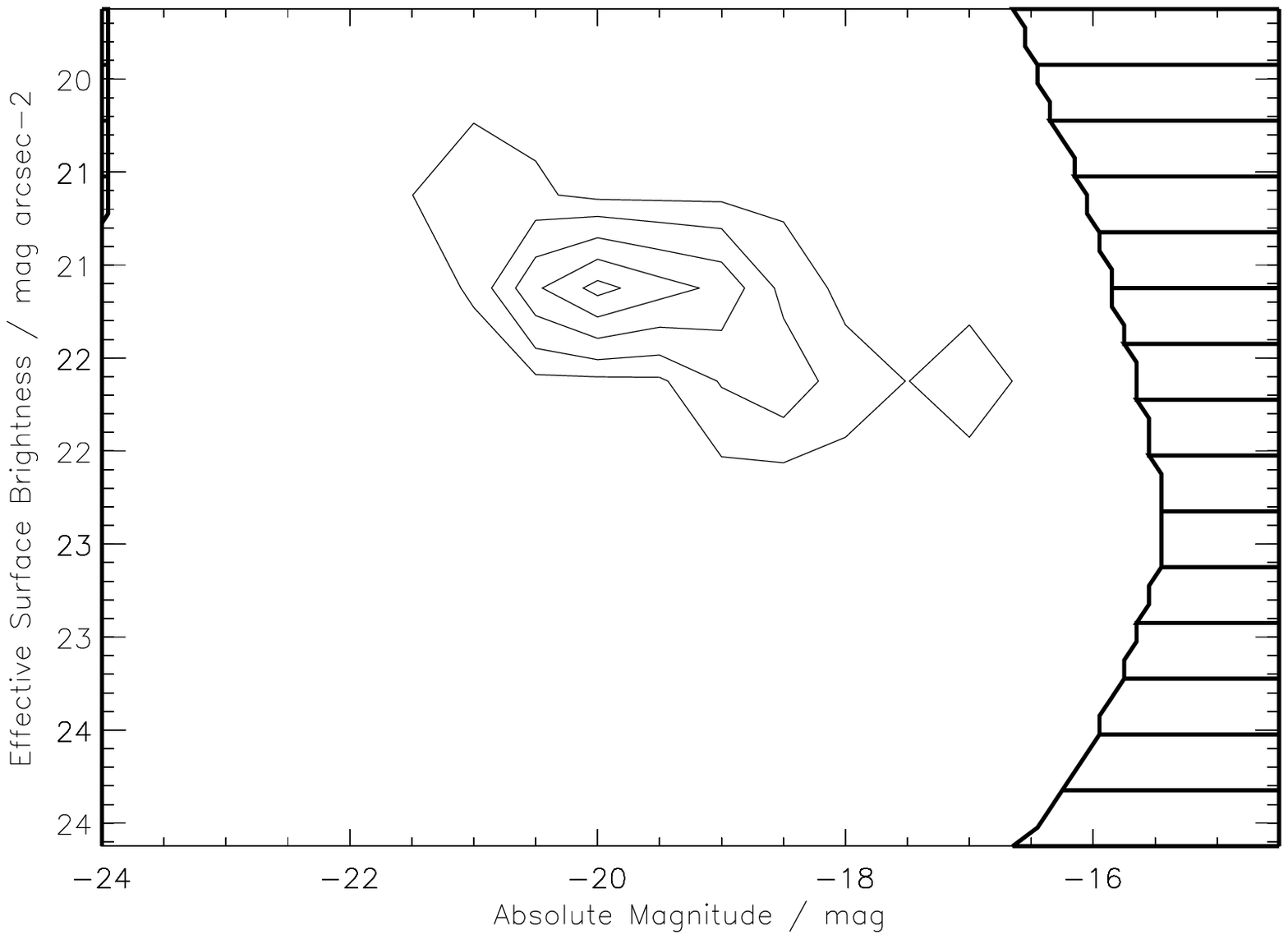,height=85.0mm,width=85.0mm}}
\caption{The BBD distribution of galaxies with $m=17.84\pm0.1$ and 
$\mu_{e}^{app}=22.31\pm0.1$. The contour lines are set at 0.01, 0.05, 0.10, 
0.15 and 0.20 chance of the galaxy being in that bin.}
\end{figure}

To generate the matrix, $I(M,\mu)$, the probability distributions for each 
of the galaxies without redshifts, (normalised to unity) are summed 
together to give the total distribution of those galaxies without redshifts. 
However these galaxies could have the full range of redshifts that each 
galaxy can be detected over, not a range limited to $z=0.12$. Therefore the 
number distribution is weighted by the fraction of 
galaxies within each bin with $z < 0.12$.

Fig. 6 shows $I(M,\mu)$ which should be compared
to Fig. 4 ($O(M,\mu)$). The distribution of Fig 6 appears broader indicating
that the missing galaxies are {\it not} random but that they are predominantly
low luminosity, low surface brightness systems. This is illustrated in
Fig. 7, which shows the ratio of $I(M,\mu)$ to $O(M,\mu)$ for the bins 
containing more than 25 galaxies with redshifts. From Fig. 7 we see that
the trend is for the ratio to increase towards the low surface 
brightness regime. There is no significant trend in absolute magnitude.
Finally we note that the although the incompleteness correction does
increase the population in some bins by as much as 50\% we shall see in \S 6 
\& 7 that the contribution from these additional systems towards the overall 
luminosity 
density is negligible.

\begin{figure}
\centerline{\psfig{file=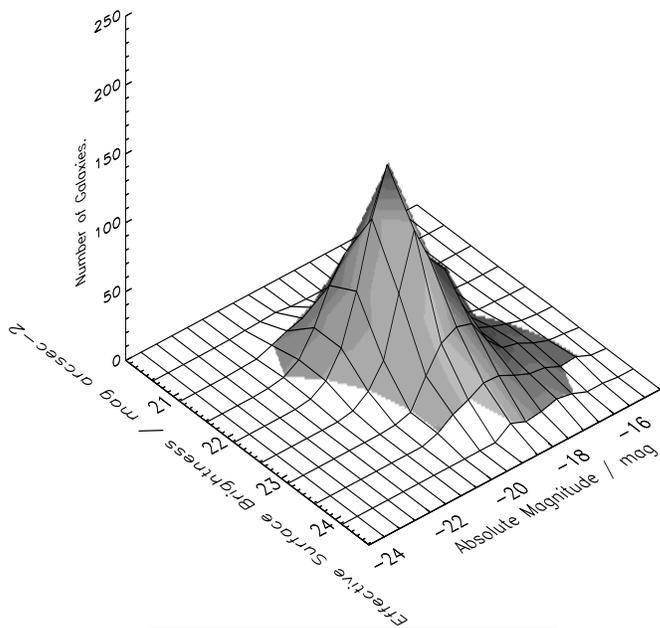,height=100.0mm,width=85.0mm}}
\centerline{\psfig{file=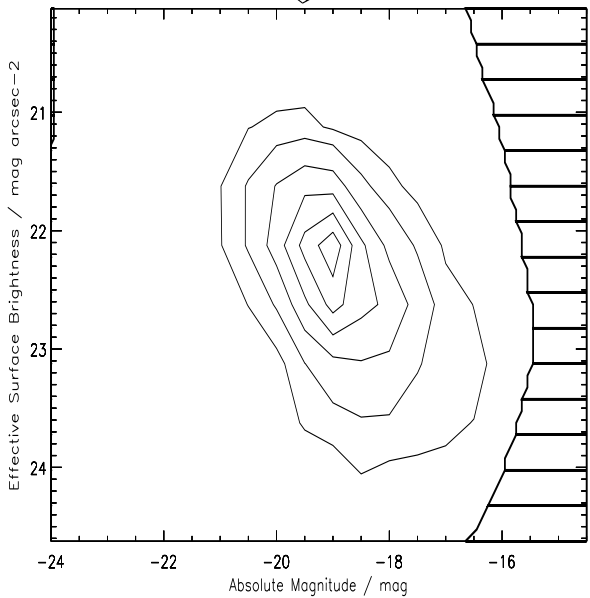,height=85.0mm,width=85.0mm}}
\caption{The bivariate number distribution of those galaxies without 
redshifts, {\it i.e.}, $I(M,\mu)$. The contour lines are set at 10, 25, 50, 
75, 100, 125 and 150 galaxies bin$^{-1}$.}
\end{figure}

\begin{figure}
\centerline{\psfig{file=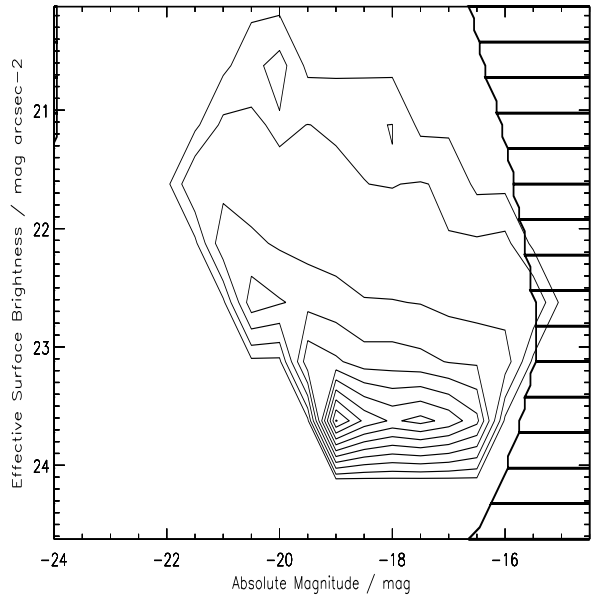,height=85.0mm,width=85.0mm}}
\caption{The ratio of galaxies without redshifts to galaxies with redshifts. 
The contour lines are set at 0.01, 0.05, 0.10, 0.15, 0.20, 0.25, 0.30, 0.35, 
0.40, 0.45 and 0.50.}
\end{figure}

\subsection{Deriving $V(M,\mu)$}
To convert the number of observed galaxies to number density per Mpc$^{3}$
a Malmquist correction is required, {\it i.e.,} $V(M,\mu)$. This matrix
reflects the volumes over which each $M$, $\mu$ bin can be observed. 
One option is to use visibility theory as prescribed by Phillipps, Davies 
\& Disney (1990) - and used to construct the constant volume line on Fig. 3.
While visibility is clearly a step in the right direction, and preferable to
applying a magnitude-only dependent correction, its limitation is that it 
assumes idealised galaxy profiles ({\it i.e.} it neglects the bulge component,
seeing, star-galaxy separation and other complications). Ideally one 
would like to extract the volume information from the data itself and this 
is possible by using a $1/V_{\rm Max}$
type prescription, {\it i.e.}, within each $O(M,\mu)$ bin, the maximum 
redshift at which a galaxy is seen is determined and the volume derived from 
this redshift. The advantages of using the data set rather than theory is that
it naturally incorporates all redshift dependent selection biases. However 
the maximum redshift is susceptible to scattering from higher-visibility bins.
An improved version is therefore to use the $90^{th}$ percentile redshift 
and to adjust $O(M,\mu)$ and $I(M,\mu)$ accordingly. 

Although this requires rejecting 10\% 
of the data it has two distinct advantages. Firstly it ensures that the
redshift distribution in each bin has a sharp cutoff (as opposed to a
distribution which peters out). Secondly it uses the entire dataset as
opposed to the maximum redshift only. In the case where the $90^{th}$
percentile is not exact we take the galaxy nearest.
Using these redshifts the volume can be calculated independently for each
bin assuming an Einstein-de Sitter cosmology as follows:

\begin{equation}
V(M,\mu)=V_{z_{min}(M,\mu)}^{z_{90}(M,\mu)}
\end{equation}

\noindent where,

\begin{equation}
V_{z_{min}(M,\mu)}^{z_{90}(M,\mu)} = \int^{z_{90}(M,\mu)}_{z_{min}(M,\mu)}\,
\frac{\sigma\,c\,d_{l}^{2}}{H_{o}\,(1+z)^{3.5}}\,dz
\end{equation}

\noindent and $\sigma$ is the solid angle in steradians on the sky. $d_l$ is 
the luminosity distance to the galaxy. $z_{min}$ is the minimum distance over 
which a galaxy can be detected. $z_{min}$ is calculated from visibility theory
(Phillipps, Davies \& Disney 1990, see also Appendix B) adopting values
for the maximum magnitude and maximum size of $m_{B}=14.00$ mag and 
$\theta=200''$ respectively.

\begin{figure}
\centerline{\psfig{file=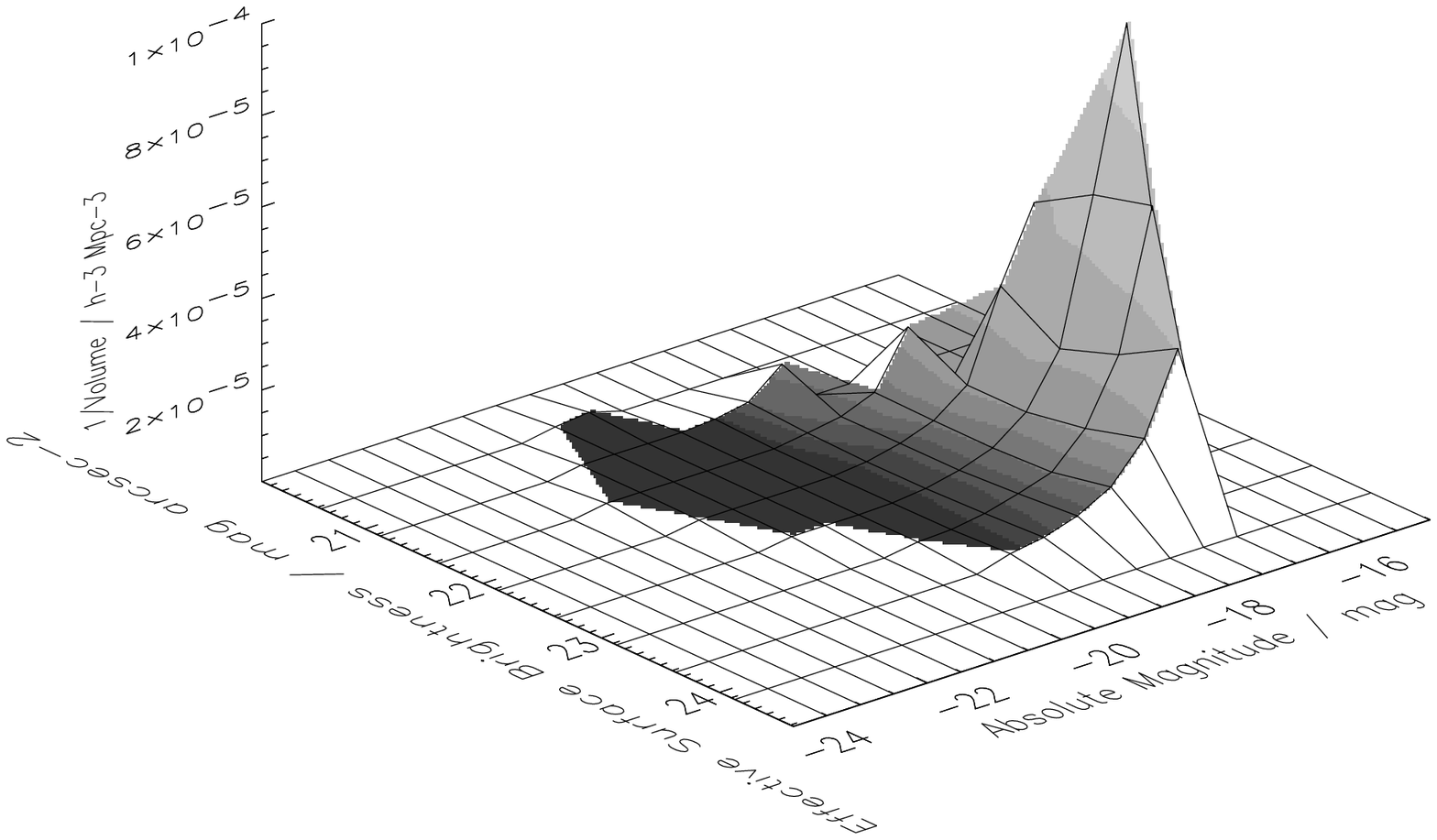,height=100.0mm,width=85.0mm}}
\centerline{\psfig{file=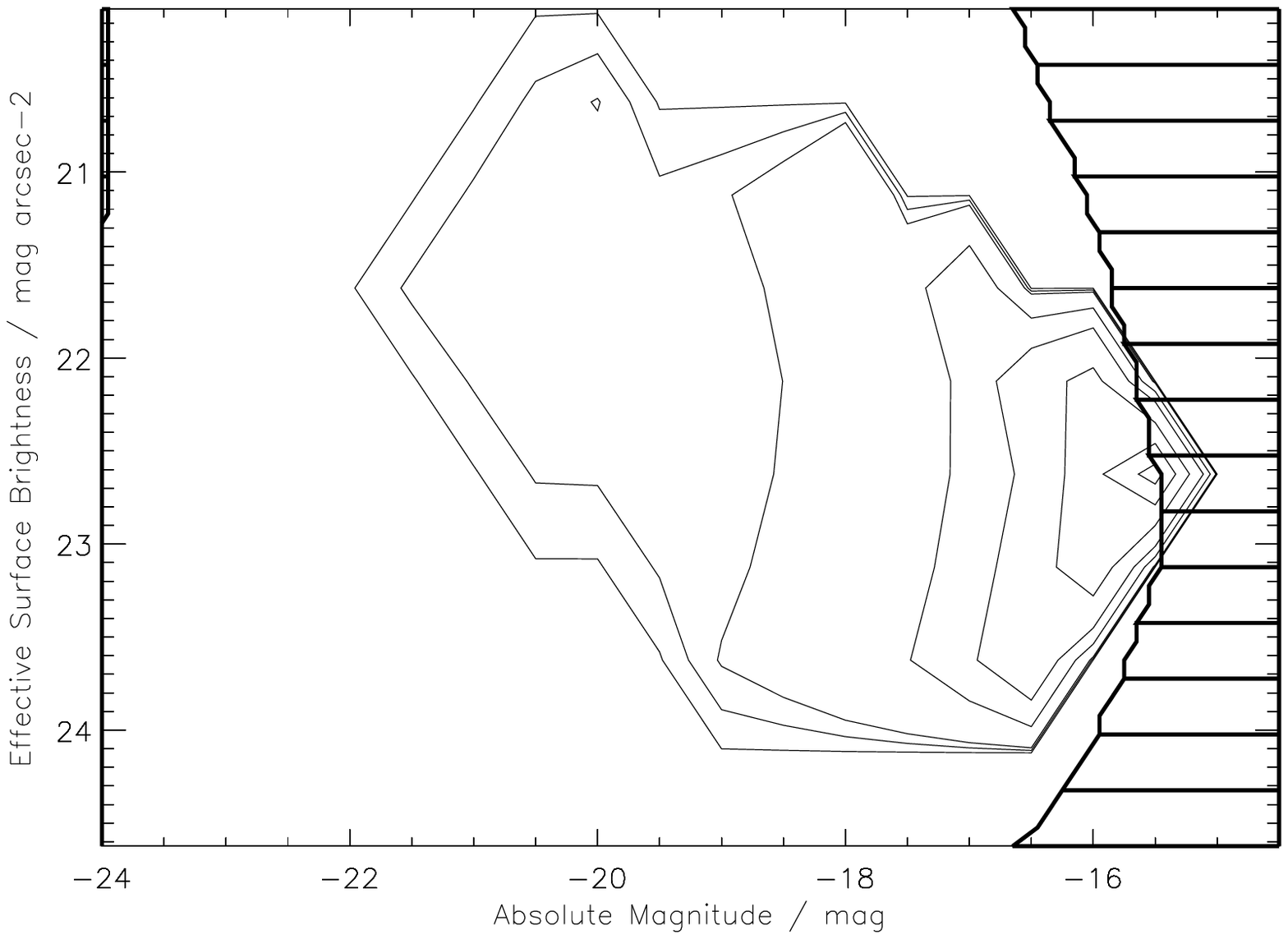,height=85.0mm,width=85.0mm}}
\caption{Plot of 1/volume in each bin. The contour lines are set 
at $1.0\times 10^{-7}$, $1.0\times 10^{-6}$, $2.0\times 10^{-6}$, 
$1.0\times 10^{-5}$, $2.0\times 10^{-5}$, $4.0\times 10^{-5}$, 
$6.0\times 10^{-5}$, $8.0\times 10^{-5}$ and $1.0\times 10^{-4}$ h$^{-3}$ 
Mpc$^{-3}$.}
\end{figure}

Fig. 8 shows the matrix $\frac{1}{V(M,\mu)}$. 
Note that squares containing fewer than 25 galaxies are not shaded.
This matrix is flat-bottomed due to the cutoff at $z=0.12$. 
Fig. 8 shows a strong dependency upon magnitude (i.e. classical
Malmquist bias as expected) and also upon surface brightness. 
This surface brightness dependency is particularly strong near the
$10^{4}$ Mpc$^{3}$ volume limit, as the data becomes sparse. Inside
this volume limit the contour lines generally mimic the curve of the 
visibility-derived volume boundary. This suggests that visibility 
theory provides a good description of the combined volume dependency.
The sharp cutoff along the high surface brightness edge may be real
but could also be a manifestation of the complex star-galaxy separation 
algorithm (see Maddox 1990a). Given that a galaxy seen over a larger distance 
appears more compact and that local dwarfs have smaller scale lengths than 
giants (cf. Mateo 1998), this seems reasonable. We will investigate this 
further through high-resolution imaging. 

The main point to take away from Fig. 8 is that the visibility
surface of the 2dFGRS input catalogue is complex and dependent on both $M$
and $\mu_e$ (although predominantly $M$). Any methodology which ignores
surface brightness information and implements a volume-bias correction in 
luminosity only, is implicitly assuming uniform visibility in surface 
brightness. The 2dFGRS data clearly show that this is not the case. 

\subsection{Clustering}
Structure is seen on the largest measurable scales (e.g. de Lapparent, Gellar
\& Huchra 1986). 
To determine whether the effects of clustering are significant we
constructed a radial density profile, as shown in Fig. 9. This was 
derived from those bins for which more than 100 galaxies are seen over the 
whole range from $0.015<z<0.12$ (i.e. $-19.75 < M < -18.75$ and $21.1 < 
\mu_e < 22.6$). Those galaxies which are brighter cannot be seen at $z < 0.015$
due to the bright magnitude cut at $m = 14.00$ and would therefore bias 
the number density towards the bright end. For 
these high-visibility galaxies we calculated their number-density ($\phi$) in 
equal volume intervals of $5.0\,\times 10^{3}$ Mpc$^{3}$, from $z=0.0185$ to 
$z=0.12$. Fig. 9 shows that clustering is severe with what appears 
to be a large local void around $z=0.04$ and walls at $z=0.06$ and $z=0.11$. 
The ESO Slice Project (ESP) survey (Zucca et al. 1997) whose line-of-sight
(RA $\sim 00$h, $\delta \sim -40$) is just outside the 2dF SGP region, 
measures an 
under-density at $\leq 140h^{-1}$Mpc ($z \approx 0.045$) and an over-density 
at $z \approx 0.1$. The structure that they see closely resembles the 
structure that we see. A reliable measure of the 
BBD needs to correct for this clustering-bias. Here we adopt a strategy
which implicitly assumes, firstly, that clustering is independent of either 
$M$ or $\mu$, and secondly, that evolutionary processes to z = 0.12 are 
negligible.

\begin{figure}
\centerline{\psfig{file=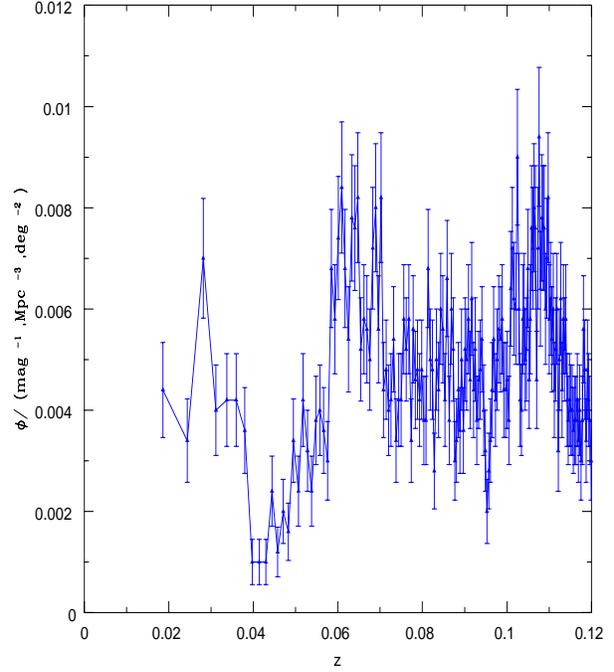,height=100.0mm,width=85.0mm}}
\caption{The clustering map. This shows the number density of giant galaxies 
($-19.75<M<-18.75$ and $21.1<\mu_e<22.6$ as a function of redshift. The points
are spaced equally in volume at intervals of 5000 h$^3$ Mpc$^3$ starting at
10000 h$^3$ Mpc$^3$.}
\end{figure}

On the basis of these caveats we constructed a weighting matrix,
$W(M,\mu)$. This was determined from the high-visibility galaxies
by taking the ratio of the number-density of high-visibility galaxies over
the full redshift range divided by the number-density of high-visibility
galaxies over the redshift range of each bin, i.e.:

\begin{equation}
W(M,\mu)=\frac {\bar{\phi}(High-vis)^{z=0.12}_{z=0.015}} 
{\phi(High-vis)^{zmax(M,\mu)}_{zmin(M,\mu)}}
\end{equation}

This weighting matrix is shown in Fig. 10. The implication is
that the number-density of low-luminosity systems will be amplified by
almost a factor of 1.5, to correct for the apparent presence of a large
local void along the SGP region - indicated by the vertical ridge at
$M = -17$ on Fig. 10. Once again this implicitly assumes that 
the clustering of dwarf and giant systems is correlated.

\begin{figure}
\centerline{\psfig{file=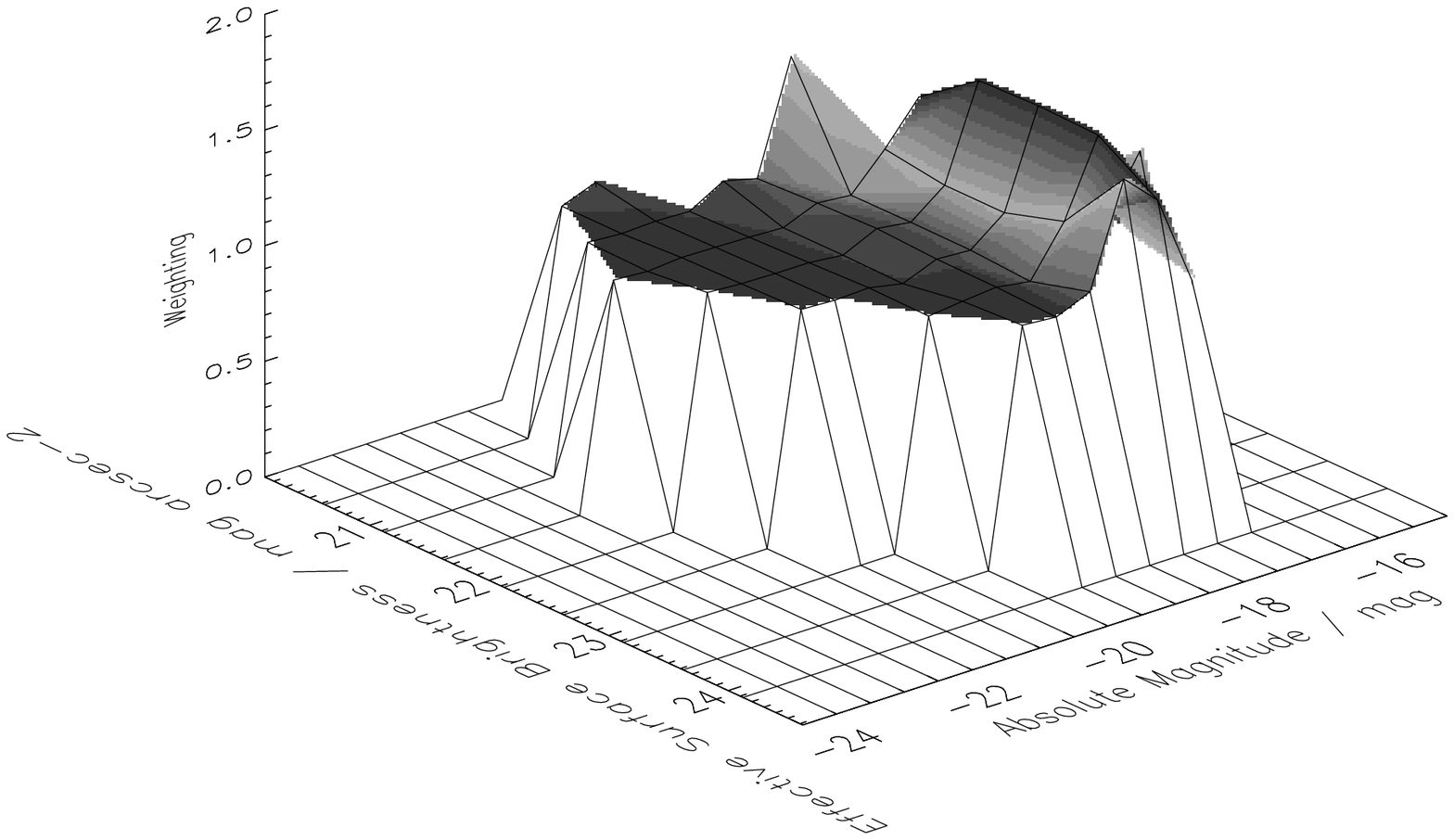,height=100.0mm,width=85.0mm}}
\centerline{\psfig{file=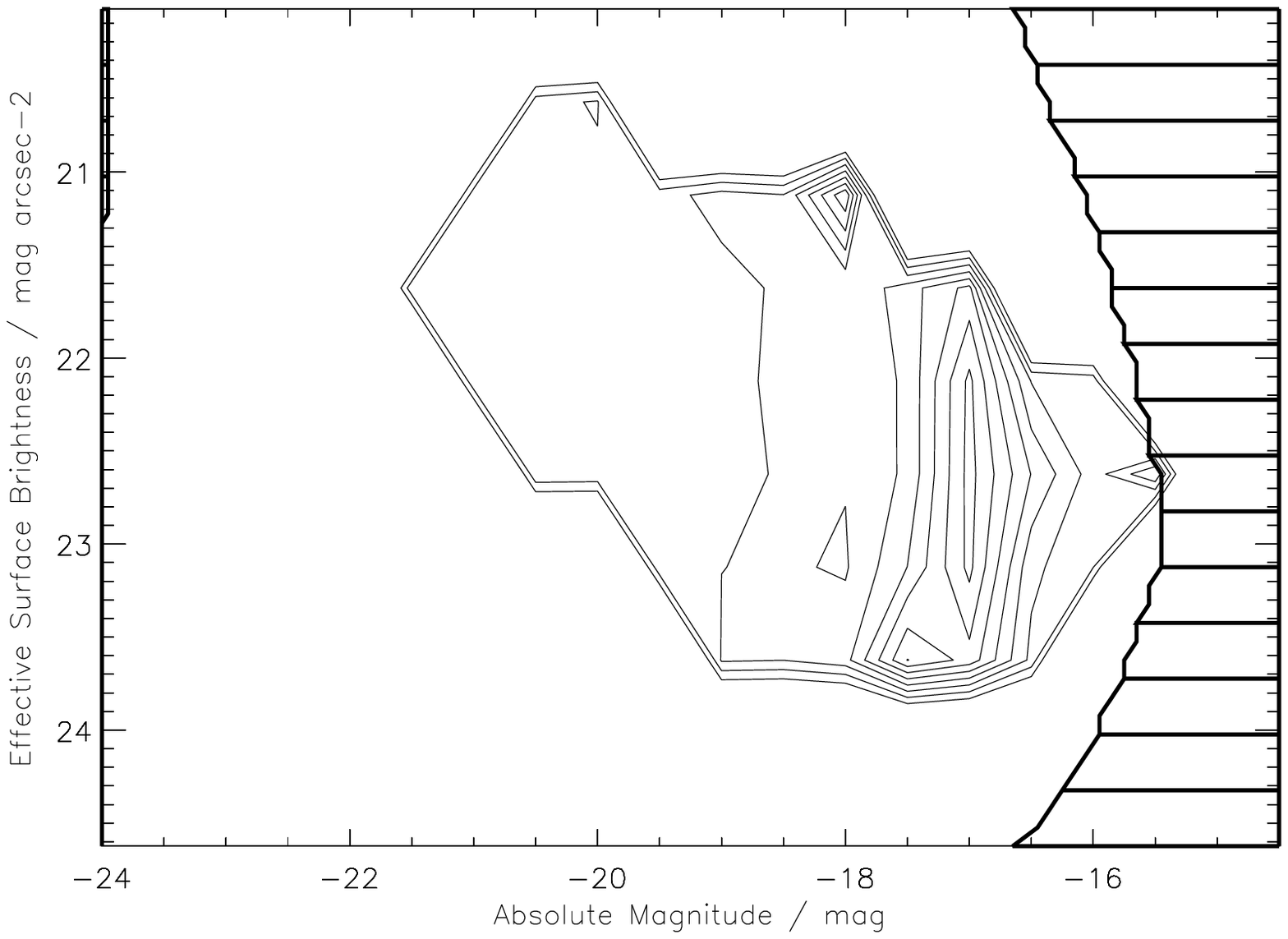,height=85.0mm,width=85.0mm}}
\caption{The Weighting Map. The contours are at 0.8, 0.9, 1.0, 1.1, 1.2, 1.3, 
1.4 and 1.5.}
\end{figure}

\section{The 2dFGRS BBD}
Finally we can combine the four matrices, $O(M,\mu)$, $I(M,\mu)$, $V(M,\mu)$
and $W(M,\mu)$ (see Eqn~\ref{eq:phiM}), to generate the 2dFGRS bivariate 
brightness distribution, as shown in Fig. 11. This depicts the underlying 
local galaxy number-density distribution inclusive of surface brightness 
selection effects. Only those bins which are based upon 25 or more galaxies 
are shown. Note that by summing the BBD along 
the surface brightness axis one recovers the luminosity distribution of 
galaxies. By summing along the magnitude axis one obtains the surface 
brightness distribution of galaxies (see \S 8).

\begin{figure}
\centerline{\psfig{file=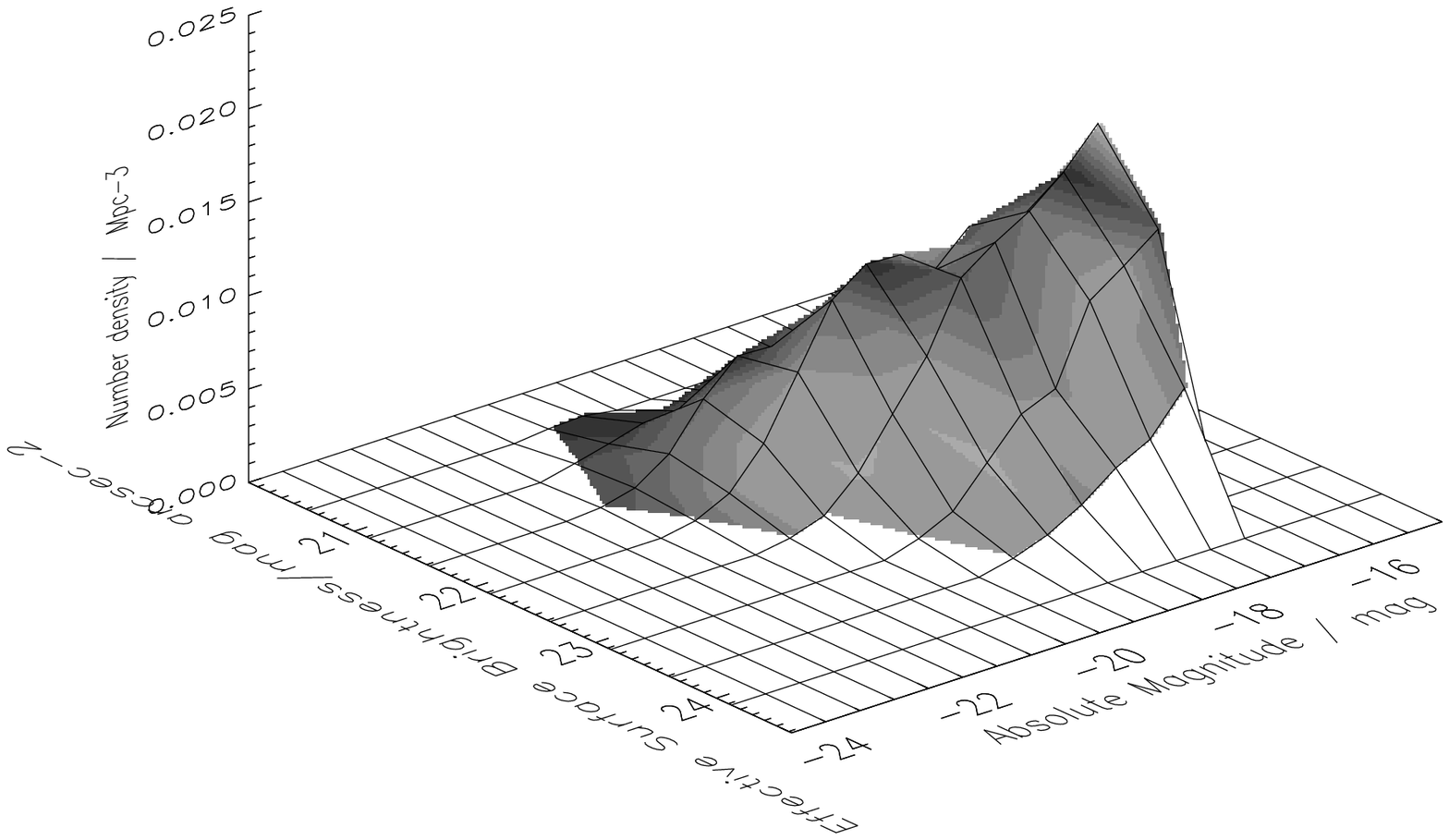,height=100.0mm,width=85.0mm}}
\centerline{\psfig{file=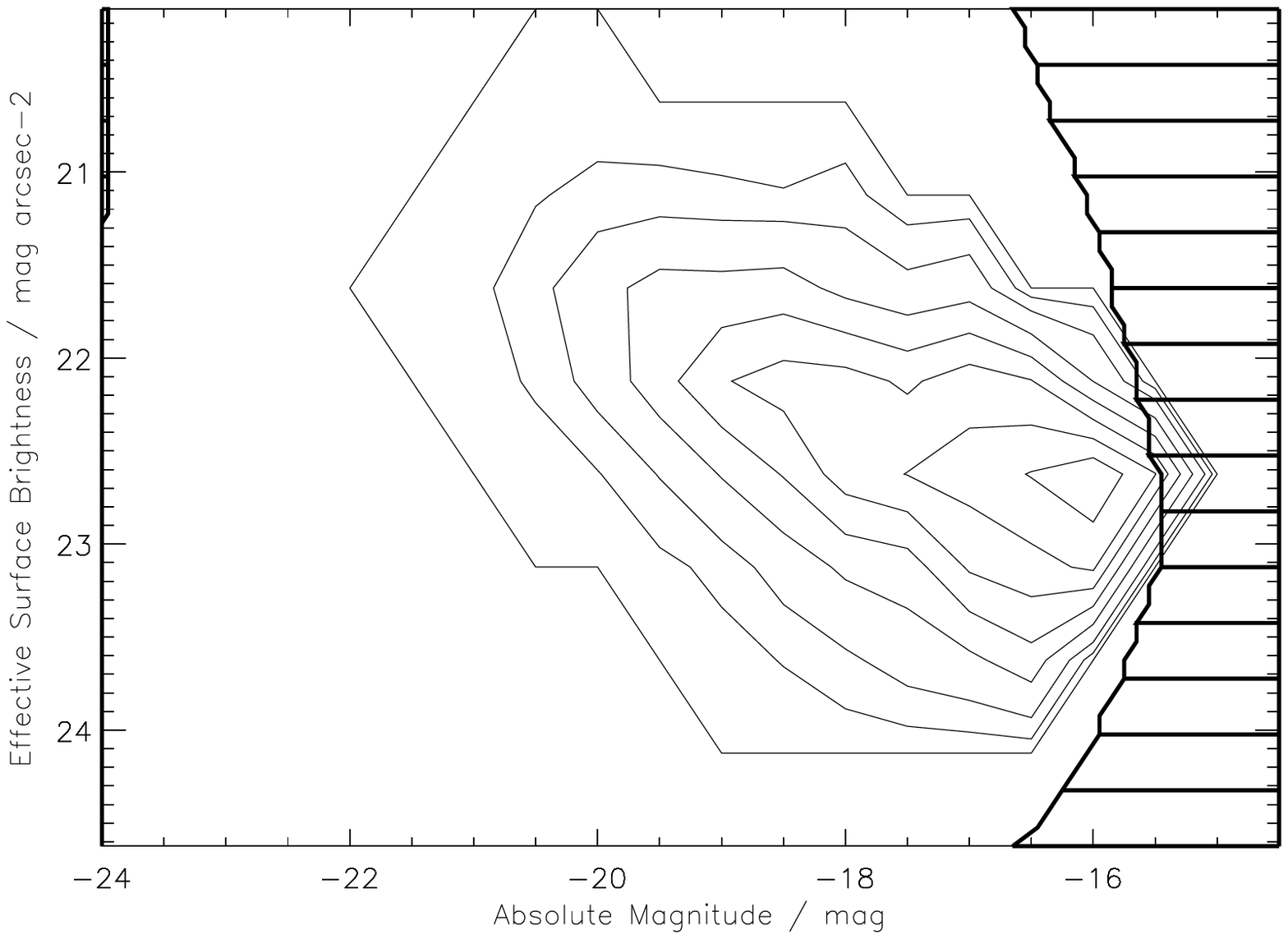,height=85.0mm,width=85.0mm}}
\caption{The 2dFGRS Bivariate Brightness Distribution.The contour lines are 
set at $1.0\times 10^{-7}$, $1.0\times 10^{-3}$, 
$2.5\times 10^{-3}$, $5.0\times 10^{-3}$, $7.5\times 10^{-3}$, 
$1.0\times 10^{-2}$, $1.25\times 10^{-2}$, $1.5\times 10^{-2}$,
$1.75\times 10^{-2}$, $2.0\times 10^{-2}$, and $2.25\times 10^{-2}$  galaxies 
Mpc$^{-3}$ bin$^{-1}$. 
The thick lines represent the selection boundaries calculated from visibility 
theory.}
\end{figure}

\begin{figure}
\centerline{\psfig{file=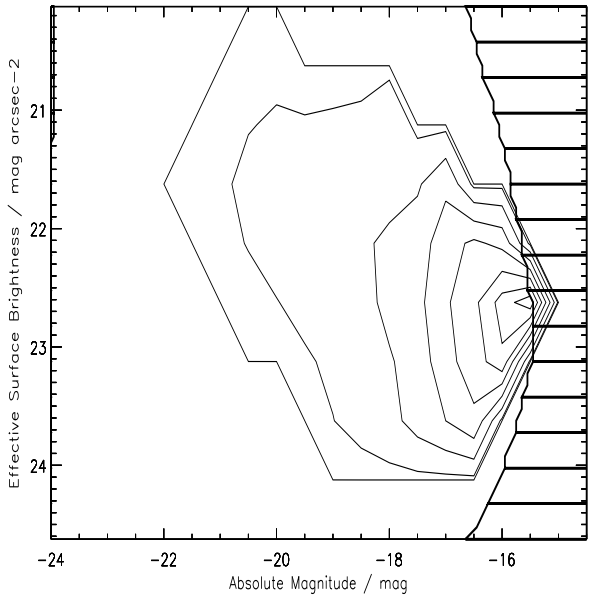,height=85.0mm,width=85.0mm}}
\caption{The errors in the BBD. The contour lines are set at 
$1.0\times 10^{-7}$, $1.0\times 10^{-4}$, $5.0\times 10^{-4}$, 
$1.0\times 10^{-3}$, $1.5\times 10^{-3}$, $2.0\times 10^{-3}$, 
$2.5\times 10^{-3}$, $3.0\times 10^{-3}$, $3.5\times 10^{-3}$ and 
$4.0\times 10^{-3}$ galaxies Mpc$^{-3}$ bin$^{-1}$.}
\end{figure}

Fig. 12 shows the errors in the BBD. These were initially determined via 
Monte-Carlo simulations, assuming a Gaussian error distribution of $\pm 0.2$ 
mag, in the APM magnitudes. This showed that the errors were proportional to
$\sqrt{(1/N)}$, and since $\sqrt{(1/N)}$ is much faster to calculate, this is 
the result that is used throughout the calculations. These errors were then 
combined in quadrature with the additional error in the volume estimate, 
assuming Poisson statistics. The total error is given by:

\begin{equation}
\sigma_{\phi}=\sqrt{(\frac{1}{N_{tot}(M,\mu)})+(\frac{1}{N_{z}(M,\mu)})}
\end{equation}

The errors become significant ($>$ 20\%), when 
$M>-16$ and around the boundaries of the BBD shown in Fig. 11. The data and 
associated errors are tabulated in Table 2. From Figs 11 \& 12 we note the 
following:

\subsection{A luminosity-surface brightness relation}
The BBD shows evidence of a luminosity-surface brightness relation similar to
that seen in Virgo, ($M_{B} \propto 1.6 \mu_{o}$, Binggeli 1993), in the 
Hubble Deep Field, ($M_{F450W} \propto 1.5\mu_e$, Driver 1999) and in Sdm 
galaxies ($M_{B} \propto 2.02\pm0.16 \mu_e$ de Jong \& Lacey 2000). 
A formal fit to the 2dFGRS data yields $M_{B} \propto (2.4 \pm ^{1.5}_{0.5}) 
\mu_{e}$. While confirming the general trend this result appears significantly 
steeper than the Virgo and HDF results. Both the Virgo and HDF results are
based on lower luminosity systems hence this might be indicative of a 
second order dependency of the relation upon luminosity. Alternatively it may
reflect slight differences in the data/analysis as neither the Virgo nor HDF 
data include isophotal corrections whereas the 2dFGRS data is more susceptible
to atmospheric seeing. The gradient is slightly steeper than the de Jong \& 
Lacey result, but is well within the errors.

The presence of a luminosity-surface brightness relation highlights concerns
over the completeness of galaxy surveys, as surveys with bright isophotal
limits will preferentially exclude dwarf systems leading to an underestimate 
of their space-densities and variations such as those seen in Fig. 1.

The confirmation of this luminosity-surface brightness relation within such
an extensive dataset is an important step forward and any credible model of 
galaxy formation must now be required to reproduce this relation.

\subsection{A Dearth of Luminous Low surface brightness galaxies}
Within each magnitude interval there appears to be a preferred range in 
surface brightness over which galaxies may exist. While the high-surface
brightness limit {\it may} be due, in part or whole, to star-galaxy separation 
and/or fibre-positioning accuracy the low surface brightness limit appears 
real. This cannot be a selection limit as one requires a mechanism which
hides luminous LSBGs yet allows dwarf galaxies of similar surface brightness 
to be detected within the same volume. The implications are that these
galaxy types (luminous-LSBGs) are rare with densities less than $10^{-4}$ 
galaxies Mpc$^{-3}$.
This result is important as it directly addresses the issues raised in the 
introduction and implies that existing surveys have {\it not} missed large
populations of luminous low surface brightness galaxies.
Perhaps more importantly it confirms that the 2dFGRS is complete for 
giant galaxies and that the postulate that the Universe might be dominated by 
luminous LSBGs (Disney 1976) is ruled out. 

One caveat however is that luminous-LSBGs could be masquerading as dwarfs.
For example consider the case of Malin 1 (Bothun {\it et al.} 1987) which
has a huge extended disk ($55$ kpc scale-length) of very low surface brightness
($\mu_o = 26.5$). This system is actually readily detectable because of its
high surface brightness active core, however within the 2dFGRS limits it would 
have been miss-classified as a dwarf system with $M=-17.9$, $\mu_e=21.8$. 
Hence Figs 11 \& 12 rule out luminous disk systems only. To determine whether 
objects such as Malin 1 are hidden amongst the dwarf population will require 
either ultra-deep CCD imaging or cross-correlation with HI-surveys which would 
exhibit very high HI mass-to-light ratios for such systems.

\subsection{The rising dwarf population}
The galaxy population shows a steady increase in number-density with
decreasing luminosity. This continues to the survey limits at $M=-16$, 
whereupon the volume limit 
and surface brightness selection effects impinge upon our
sample. The expectation is that the distribution continues to rise and
hence the location of the peak in the number-density distribution remains 
unknown. However we do note that the increase seen within our selection limits
is insufficient for the dwarf population to dominate the luminosity-density as 
shown in the next section. 

Perhaps more surprising is the lack of sub-structure indicating either a 
continuity between the giant and dwarf populations or that any sub-structure is
erased by the random errors. The former case is strongly indicative of a
hierarchical merger scenario for galaxy formation which one expects to lead
towards a smooth number-density distribution between the dwarf and giant 
systems (White \& Rees, 1978). This is 
contrary to the change in the luminosity distribution
of galaxies seen in cluster environments (e.g. Smith, Driver, \& Phillipps 
1997). In a later paper we intend to explore the dependency of the 
BBD upon environment.

\section{The value of $j_{B}$ and $\Omega_{M}$}
The luminosity density matrix, $j(M,\mu)$ is constructed from $L\,\Phi(M,\mu)$
in units of $L_{\odot}$Mpc$^{-3}$ and is shown as Fig. 13. The distribution
is strongly peaked close to the conventional $M_{*}$ parameter derived
in previous surveys (see Table 1). The peak lies at $M = -19.5 $ mag and 
$\mu_e = 22.12$ mag arcsec$^{-2}$. The final value obtained  
is $j_B=(2.49 \pm 0.20) \times 10^{8} h_{100} L_{\odot}$Mpc$^{-3}$. The sharp 
peak sits firmly in the centre of our observable region of the 2dFGRS BBD and 
drops rapidly off on all sides towards the 2dFGRS BBD boundaries. This 
implies that while the 2dFGRS does not survey the entire parameter space 
of the known BBD it does effectively contain the full galaxy 
contribution to the local luminosity density. Redoing the calculations using
galaxies with redshifts also results in $j_B=(2.49 \pm 0.20) \times 10^{8} 
h_{100} L_{\odot}$Mpc$^{-3}$. This demonstrates that there is no dependency of
the results upon the assumption made for the distribution of galaxies without 
redshifts. 

We note that $j$ 
derived via a direct $\frac{1}{V_{max}}$ estimate, without any surface 
brightness or clustering
corrections, (Eqn~\ref{eq:j}) gives a value of $j=1.82\pm0.07\times 
10^{8} h_{100} L_{\odot}$Mpc$^{-3}$. Including the isophotal magnitude 
correction only leads to a value of $j=2.28\pm0.09\times10^{8} h_{100} 
L_{\odot}$Mpc$^{-3}$. Hence a more detailed analysis leads
to a 36.8\% increase in $j$, of this 36.8\%, 25.3\% is due to the isophotal
correction, 10.4\% is due to the Malmquist bias correction and 1.1\% is due 
to the clustering correction.

The final value 
agrees well with that obtained from the recent ESO Slice Project (Zucca 
{\it et al.} 1997). The method that they used corrects for clustering, but 
not surface brightness --- although their photometry is based on aperture 
rather than isophotal magnitudes. However it is worth pointing out that $j$ 
is sensitive to the exact value of $\mu_{lim}$. The quoted error in 
$\mu_{lim}$ is $\pm 0.3$ (for plate-to-plate variations, Metcalfe et al 1995; 
Pimblett et al 2000). Table 3 shows a 
summary of results when repeating the entire analysis using the upper and 
lower error limits. It therefore seems likely that a {\it combination} of
surface brightness biases and large scale structure can indeed lead to the 
type of variations seen in Table 1.

\begin{figure}
\centerline{\psfig{file=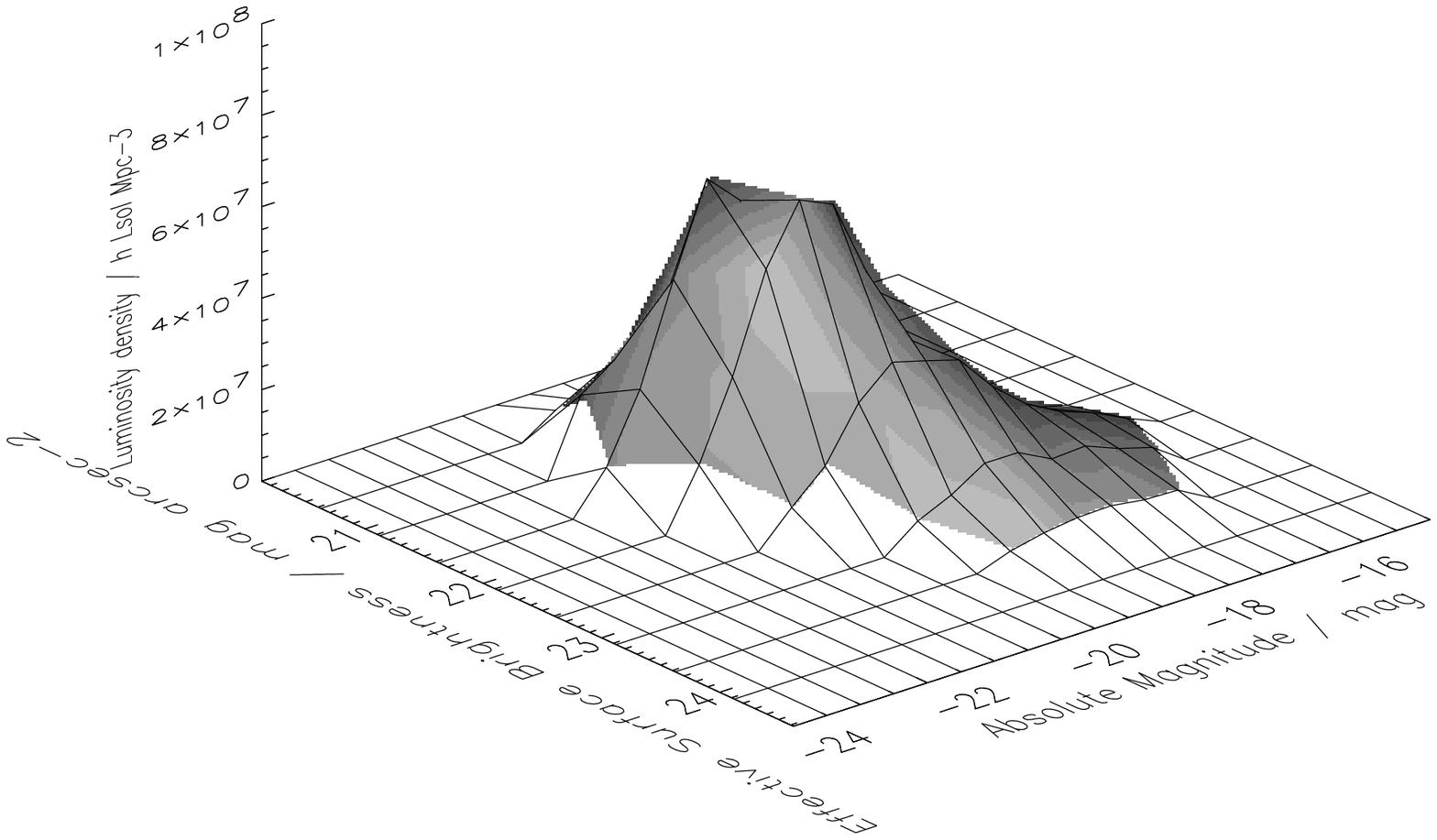,height=100.0mm,width=85.0mm}}
\centerline{\psfig{file=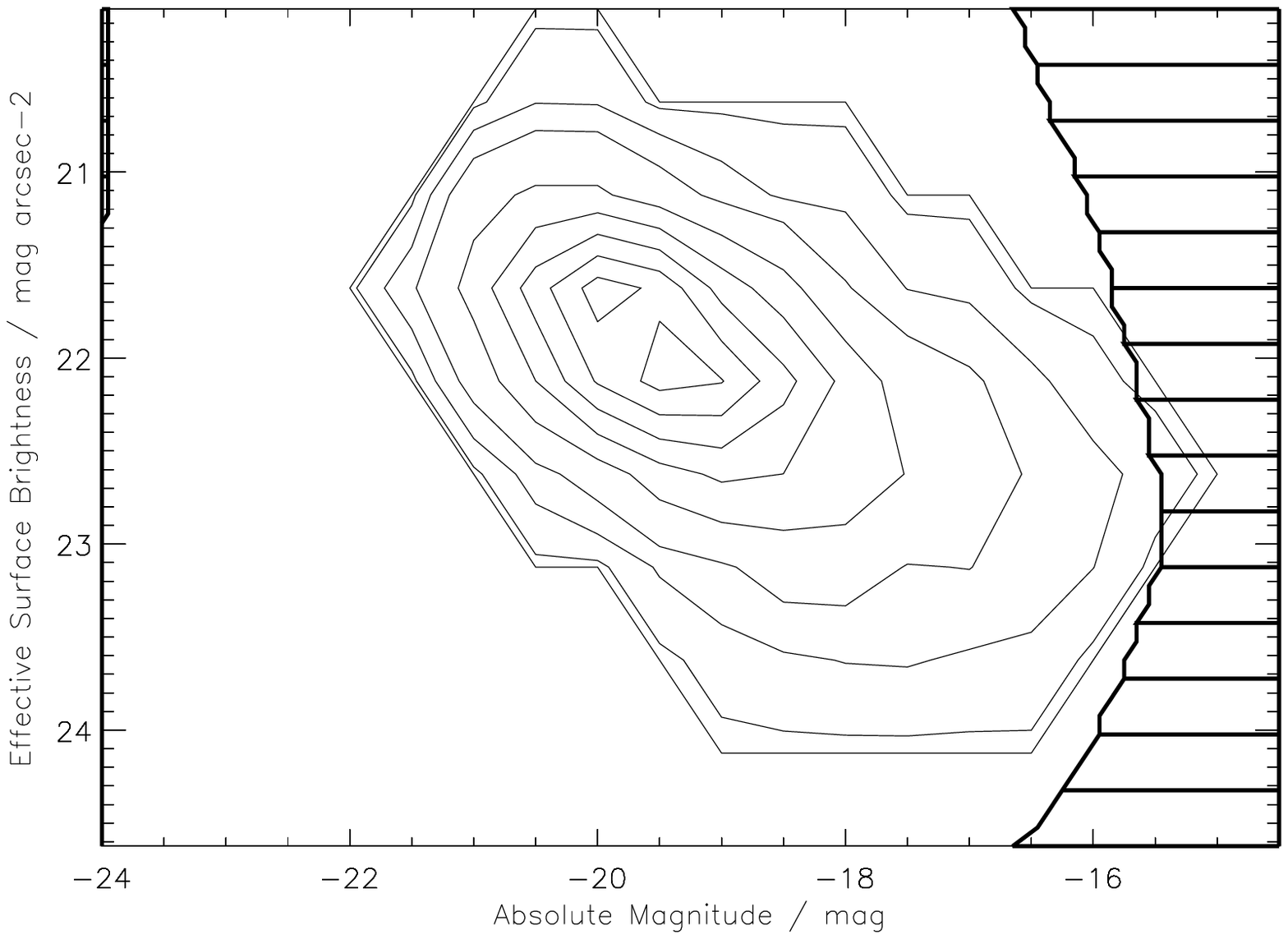,height=85.0mm,width=85.0mm}}
\caption{Plot of the Luminosity Density Distribution. The contour lines are 
set at 100, $1.0\times 10^6$, $5.0\times 10^6$, $1.0\times 10^7$, 
$2.0\times 10^7$, $3.0\times 10^7$, $4.0\times 10^7$, $5.0\times 10^7$, 
and $6.0\times 10^7$L$_{\odot}$ Mpc$^{-3}$ bin$^{-1}$.}
\end{figure}

Finally following the method of Carlberg, Yee \& Ellingson (1997) we
can obtain a crude ball-park figure for the total local mass density by 
adopting a universal mass-to-light ratio based on that observed in clusters. 
While this method neglects biasing (White, Tully \& Davis, 1988) it does 
provide a useful crude {\it upper} limit to the mass density. From Carlberg 
{\it et al.} (1997) we find: $\frac{{\cal M}_{Dyn}}{L_R} = 289\pm50 h_{100} 
\frac{{\cal M}_{\odot}}{L_{\odot}}$. Assuming a mean colour of $(B-R)=1.1$  
and a solar colour index of 1.17 this converts to: 
$\frac{{\cal M}_{Dyn}}{L_B} = 271\pm47 h_{100}
\frac{{\cal M}_{\odot}}{L_{\odot}}$. Multiplying the luminosity density by the 
mass-to-light ratio yields a value for the local mass-density of 
$\Omega_{M} \approx 0.24 $. We note that this is consistent with
the current constraints from the combination of Sn Ia results with the recent 
Boomerang and Maxima-1 results (Balbi {\it et al.} 2000; de Bernardis 
{\it et al.} 2000).

\section{Comparisons with other surveys}
As this work represents the first detailed measure of the field BBD there is
no previous work with which to compare. However as mentioned earlier it is
trivial to convert the BBD into either a luminosity distribution and/or a
surface brightness distribution both of which have been determined by 
numerous groups. This is achieved by summing across either luminosity or 
surface brightness intervals. For those bins containing fewer than 25 galaxies
for which a volume- and clustering- bias correction were not obtained we
use the volume-bias correction from the nearest bin with 25 or more galaxies. 
This will lead to a slight underestimate in the number-densities, however as 
the number-density peak is well defined this effect is negligible.

\subsection{The luminosity distribution}
Fig. 14 shows a compendium of luminosity function measures (see Table 1 and 
Fig. 1). Superimposed on the previous Schechter function fits (dotted lines)
are the results from the 2dFGRS.
Three results are shown, the luminosity distribution neglecting surface 
brightness and clustering (short dashed line), the luminosity distribution
inclusive of surface brightness corrections (long dashed line) and finally 
the luminosity distribution inclusive of surface brightness and clustering 
corrections (solid line). The inset shows the $2\sigma \,\chi^{2}$ error
eclipse for the latter case, yielding Schechter function parameters of
$M_{*} = -19.75\pm0.05, \alpha=-1.09\pm0.03$ and $\phi_{*}=0.0202\pm0.0002$ in
close agreement with the ESO Slice Project (ESP, see Table 1).

Comparing the 2dFGRS result with  other surveys suggests that the main effect
of the surface brightness correction is to shift $M_{*}$ brightwards by the 
mean isophotal correction of 0.33 mag and to increase the number density 
by a factor of 1.2. The clustering correction has little effect at bright 
magnitudes, almost by definition, but significantly amplifies the dwarf 
population. This is a direct consequence of an apparent local void at z=0.04 
along the full range of the 2dF SGP region. 

We note that the combination of these two corrections mimics closely
the discrepancies between various Schechter function fits. For example the
shallower SSRS2, APM and Durham/UKST surveys are biased at all magnitudes by
the apparent local void while the deeper ESP and Autofib surveys (which employ 
clustering independent methods) probe beyond the void.
Similarly because of the luminosity-surface brightness relation those surveys
which probe to lower surface brightnesses will find a higher dwarf-to-giant 
ratio.

The shaded region shows the point at which our data starts to become highly
uncertain because of the small volume surveyed ($V < 10,000$ Mpc$^{3}$).
{\it As this is the first statistically significant investigation into the 
bivariate brightness distribution we can conclude that as yet no survey 
contains any direct census of the space density of $M_{B} > -16.0$ mag 
galaxies.}

\begin{figure} 
\centerline{\psfig{file=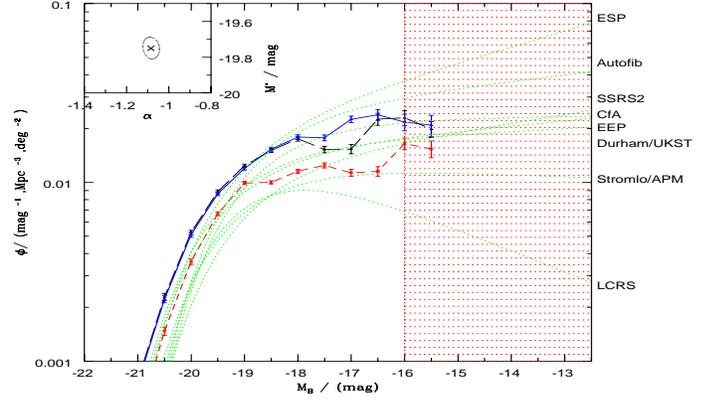,height=60.0mm,width=85.0mm}}
\caption{The solid line shows the final luminosity function, with all 
corrections taken into account. The long dashed line shows the luminosity 
function with surface brightness corrections only, and the short dashed line 
shows the luminosity function ignoring surface brightness and clustering 
corrections. Also shown, as 
dotted lines, are the luminosity functions from Fig. 1. The shaded region 
denotes the limit of reliability based on visibility theory.}
\end{figure}

\subsection{The surface brightness distribution}
There have been a few Surface Brightness Functions published over the 
years. The first was the Freeman (1970) result, which showed a Gaussian
distribution with $\bar{\mu}=21.65$ and $\sigma_{\mu}=0.3$. Since then, 
however, many galaxies have been found at greater than 10$\sigma$ from the 
mean. The probability of a galaxy occurring at 10$\sigma$ or greater is 
$\approx 1 \times10^{-20}$. The total number of galaxies in the universe is 
in the range of $10^{11}$ --- 
$10^{12}$, so the Freeman SBF must be underestimating the LSBGs. The 
distribution seen by Freeman is almost certainly due to the relatively bright 
isophotal detection threshold that the observations were taken at, around 22 -
23 mag arcsec$^{-2}$.  A more recent measure of the SBF comes from
O'Neil \& Bothun (2000) who also show a compendium of results from other
groups. The O'Neil \& Bothun data in contrast to the Freeman result
shows a flat distribution over the range $22<\mu_o<25$ albeit with substantial
scatter. Fig. 15 reproduces Fig. 1. of O'Neil \& Bothun but now includes the
final 2dFGRS results. In order to compare our results directly with the
O'Neil \& Bothun data we assumed a mean bulge-to-total ratio of 0.4 
(see Kent 1985) resulting in a uniform offset of $0.55$ mag arcsec$^{-2}$.

The 2dFGRS data is substantially broader than the Freeman distribution and 
appears to agree well with the compendium of data summarised in 
O'Neil \& Bothun (2000). From visibility theory 
(see Fig. 3) we are complete ({\it i.e.} the volume observed is greater than 
$10^{4}$ Mpc$^{3}$ and therefore statistically representative) in central 
surface brightness from $18.0 < \mu_{o} < 23$ for $M < -16$. 
Assuming that the luminosity-surface brightness relation continues
as reported in Driver (1999) and Driver \& Cross (2000) the expectation is
that the surface brightness distribution will steepen as galaxies with lower
luminosities are included.

\begin{figure} 
\centerline{\psfig{file=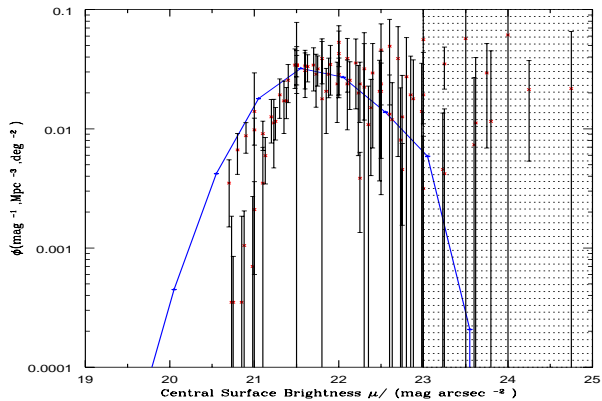,height=60.0mm,width=85.0mm}}
\caption{The solid line shows the final surface brightness function, with all
the surface brightness and clustering corrections. The points are offset 
towards the low surface brightness direction by 0.55 mag to crudely correct
for the bulge. Also shown is the O'Neil \& Bothun (2000) data, as crosses with
error bars. The shaded region is where the visibility line shown in the 
previous diagrams starts to cross the data and so the data is no longer 
completely reliable.}
\end{figure}

\section{Conclusions}
We have introduced the bivariate brightness distribution as a means
by which the effect of surface brightness selection biases in large 
galaxy catalogues can be investigated.
By correcting for the light below the isophote and including a surface 
brightness dependent Malmquist correction
we find that the measurement of the luminosity density is increased by 
$\sim 37$\% over the traditional $\frac{1}{V_{max}}$ method of evaluation. 
The majority (25\%) of this increase comes from the isophotal correction with 
10\% due to incorporating a surface brightness dependent Malmquist correction and 1\% due to the clustering correction.

We have shown that our isophotal correction is suitable for all
galaxy types and that isophotal magnitudes without correction severely 
underestimate the magnitudes for specific galaxy types.
We also note that the redshift incompleteness suggests that predominantly
low surface brightness galaxies are being missed however we also show that
these systems are predominantly low luminosity and hence contribute little to
the overall luminosity density. This is in part due to the high completeness
of the 2dFGRS ($\sim 91$ \%).

We rule out the possibility of the 2dFGRS missing a significant population 
of luminous giant galaxies down to $\mu_{e} = 24.5$ mag arcsec$^{-2}$ 
(or $\mu_{o}=23.5$ mag arcsec$^{-2}$) and note that the contribution at
surface brightness limits below $\mu_{e}=23$ mag arcsec$^{-2}$ is small
and declining. Dwarf galaxies greatly outnumber the giants and the
peak in the number density occurs at the low luminosity selection boundary.
The implication is that the most numerous galaxy type lies at $M > -16.0$
mag. The galaxy population as a whole follows a luminosity-surface brightness
relation ($M \propto (2.4 \pm ^{1.5}_{0.5}) \mu_{e}$), similar but slightly 
steeper, to that seen 
in Virgo, in the Hubble Deep Field and in SdM galaxies. This relation 
provides an additional constraint which galaxy formation models must satisfy.

We conclude that our measure of the galaxy contribution to the 
luminosity density is robust and dominated by conventional 
giant galaxies with only a small ($<10$\%) contribution from dwarf 
($M > -17.5$ mag) and/or low surface brightness giants ($\mu_e > 23$ mag 
arcsec$^{-2}$) within the selection boundaries($-24 < M_{B} < -15.5$ mag, 
$18.0 < \mu_e < 24.5$ mag arcsec$^{-2}$). However we cannot rule out the 
possibility of a contribution from an independent population outside of our 
optical selection boundaries.

Our measurement of the luminosity density is $j = 2.49\pm0.20\times10^{8} 
h_{100}$L$_{\odot}$Mpc$^{-3}$ and using a typical cluster mass-to-light ratio
leads to an estimate of the matter density of order $\Omega_{M} \sim 0.24$
in agreement with more robust measures. 

Finally we note that the bivariate brightness distribution offers a means of 
studying the galaxy population and luminosity-density as a 
function of environment and epoch fully inclusive of surface brightness 
selection biases. Future extensions to this work will include the measurement 
of BBDs and ``population peaks'' for individual spectral/morphological types,
and as a function of redshift and environment.
This step forward has only become possible because of the recent availability
of large redshift survey databases such as that provided by the 2dFGRS.

\section*{Acknowledgments}
The data shown here was obtained via the two-degree field facility on the 
3.9m Anglo-Australian Observatory. We thank all those involved in the
smooth running and continued success of the 2dF and the AAO.

\section*{References}

\begin{description}

\item{\hspace{-0.75cm}} Babul, A., Rees, M., 1992, MNRAS, 255, 346

\item{\hspace{-0.75cm}} Babul, A., Ferguson, H.C., 1996, ApJ, 458, 100

\item{\hspace{-0.75cm}} Balbi A., Ade P., Bock J., Borrill J., Boscaleri A., 
de Bernardis P., Ferreira P.G., Hanany S., Hristov V.V., Jaffe A.H., Lee A.T.,
Oh S., Pascale E., Rabii B., Richards P.L., Smoot G.F., Stompor R., Winant C.D., Wu J.H.P. 2000, ApJL in press (astro-ph 0005124)

\item{\hspace{-0.75cm}} Beijersbergen, M., de Blok, W.J.G., van der Hulst, J.M.
1999, A\&A, 351, 903

\item{\hspace{-0.75cm}} Binggeli B., 1993 in ESO/OHP Workshop on Dwarf 
Galaxies, Eds Meylan G., \& Prugniel, P., (Publ: ESO, Garching), 13

\item{\hspace{-0.75cm}} Binggeli, B., Sandage, A., Tammann, G.A., 1988, ARA\&A, 26, 509

\item{\hspace{-0.75cm}} Bothun, G.D., Impey, C.D., Malin, D.F., Mould, J.R. 1987, 
AJ, 94, 23

\item{\hspace{-0.75cm}} Boyce, P.J., Phillipps, S., 1995, A\&A, 296, 26

\item{\hspace{-0.75cm}} Bristow, P.D., Phillipps, S., 1994, MNRAS, 267, 13

\item{\hspace{-0.75cm}} Carlberg, R.G., Yee, H.K.C., Ellingson, E., Abraham, R. Gravel, P., 
Morris, S. Pritchet, C.J. 1996, ApJ, 462, 32

\item{\hspace{-0.75cm}} Carlberg, R.G., Yee, H.K.C., Ellingson, E. 1997, ApJ, 
478, 462

\item{\hspace{-0.75cm}} Colless, M. 1999, Phil.Trans.RSoc., 357, 105

\item{\hspace{-0.75cm}} Dalcanton, J.J., Spergel, D.N., Gunn, J.E., Schmidt, M., Schneider, D.P., 1997, AJ, 114, 635

\item{\hspace{-0.75cm}} Dalcanton, J.J., 1998, ApJ, 495, 251

\item{\hspace{-0.75cm}} de Bernardis, P., Ade P.A.R., Bock J.J., Bond J.R., 
Borrill J., Boscaleri A., Coble K., Crill B.P., De Gasperis G., Farese P.C.,
Ferreira P.G., Ganga K., Giacometti M., Hivon E., Hristov V.V., Iacoangeli A.,
Jaffe A.H., Lange A.E., Martinis L., Masi S., Mason P., Mauskopf P.D., 
Melchiorri A., Miglio L., Montroy T., Netterfield C.B., Pascale E., Piacentini
F., Pogosyan D., Prunet S., Rao S., Romeo G., Ruhl J.E., Scaramuzzi F., 
Sforna D., Vittorio N. 2000, Nature, 404, 955

\item{\hspace{-0.75cm}} de Jong R. \& Lacey C. 2000. Accepted for publication 
in ApJ

\item{\hspace{-0.75cm}} de Lapparent, V., Gellar, M.J., Huchra, J.P. 1986, ApJ, 304, 585

\item{\hspace{-0.75cm}} de Vaucouleurs, G. 1948, Ann. Astrophys., 11, 247

\item{\hspace{-0.75cm}} Disney, M. 1976, Nature, 263, 573

\item{\hspace{-0.75cm}} Driver, S.P., Phillips, S., Davies, J.I., Morgan, I., Disney, M.J. 1994, MNRAS, 266, 155

\item{\hspace{-0.75cm}} Driver, S.P., Windhorst, R.A., Griffiths, R.E. 1995, ApJ, 453, 48

\item{\hspace{-0.75cm}} Driver, S.P., Couch, W.J., Phillipps, S., 1998, MNRAS, 301, 369

\item{\hspace{-0.75cm}} Driver, S.P. 1999, ApJL, 526, 69

\item{\hspace{-0.75cm}} Driver, S.P., Cross N.J.G. 2000, 
in {\it Mapping the Hidden Universe}, Eds R. Kraan-Korteweg, P. Henning \& 
H. Andernach (Publ: Kluwer), (astro-ph 0004201)

\item{\hspace{-0.75cm}} Efstathiou, G., Ellis, R., Peterson, B. 1988, MNRAS, 232, 431

\item{\hspace{-0.75cm}} Ellis, R.S., Colless, M., Broadhurst, T., Heyl, J., Glazebrook, K. 
1996, MNRAS, 280, 235

\item{\hspace{-0.75cm}} Ellis, R.S., 1997, ARA\&A, 35, 389

\item{\hspace{-0.75cm}} Felten, J.E., 1985, Com Ap, 11, 53

\item{\hspace{-0.75cm}} Ferguson, H.C., Binggeli, B. 1994, A\&ARv, 6, 67

\item{\hspace{-0.75cm}} Fern\'andez-Soto A., Lanzetta K., Yahil A., 1998, ApJ, 513, 34

\item{\hspace{-0.75cm}} Folkes, S., Ronen, S., Price, I., Lahav, O., Colless, M., Maddox, S., 
Deeley, K., Glazebrook, K., Bland-Hawthorn, J., Cannon, R., Cole, S., 
Collins, C., Couch, W., Driver, S., Dalton, G., Efstathiou, G., Ellis, R., 
Frenk, C., Kaiser, N., Lewis, I., Lumsden, S., Peacock, J., Peterson, B., 
Sutherland, W., Taylor, K. 1999, MNRAS, 308, 459

\item{\hspace{-0.75cm}} Freeman, K. 1970, ApJ, 160, 811

\item{\hspace{-0.75cm}} Fukugita M., Hogan C.J., Peebles P.J.E., 1998, ApJ, 503, 518

\item{\hspace{-0.75cm}} Gaztanaga, E., Dalton, G.B. 2000, MNRAS, 312, 417

\item{\hspace{-0.75cm}} Impey, C., Bothun, G. 1997, ARA\&A, 35, 267

\item{\hspace{-0.75cm}} Jerjen, H. Binggeli, B. 1997, 'The Nature of Elliptical
Galaxies', Eds. M. Arnaboldi, G.S. Da Costa \& P. Saha, (Canberra: Mt Stromlo
Observatory) p. 239

\item{\hspace{-0.75cm}} Kent, S.M. 1985, ApJS, 59, 115

\item{\hspace{-0.75cm}} Koo D.C., Kron R.G., 1992, ARA\&A, 30, 613

\item{\hspace{-0.75cm}} Lilly, S. J., Le Fevre, O., Hammer, F., Crampton, D. 1996, ApJ, 460, 1

\item{\hspace{-0.75cm}} Lin, H., Kirshner, R., Shectman, S., Landy, S., Oemler, A., 
 Tucker, D., Schechter, P. 1996, ApJ, 464, 60

\item{\hspace{-0.75cm}} Loveday, J., Peterson B. A., Efstathiou, G., Maddox, S.
J. 1992, ApJ, 390, 338

\item{\hspace{-0.75cm}} Loveday, J., 1997, ApJ, 489, 29

\item{\hspace{-0.75cm}} Loveday, J., 2000, MNRAS, 312, 557

\item{\hspace{-0.75cm}} Madau, P., Della Valle, M., Panagia, N. 1998, ApJ, 498, 106

\item{\hspace{-0.75cm}} Maddox, S.J., Sutherland, W.J., Efstathiou, G., Loveday, J.
1990, MNRAS, 243, 692

\item{\hspace{-0.75cm}} Maddox, S.J.,  Efstathiou, G., Sutherland, W.J.
1990, MNRAS, 246, 433

\item{\hspace{-0.75cm}} Margon, B. 1999, Phil.Trans.RSoc., 357, 105

\item{\hspace{-0.75cm}} Marzke, R., Huchra, J., Geller M. 1994, ApJ, 428, 43

\item{\hspace{-0.75cm}} Marzke, R., Da Costa, N., Pelligrini, P., Willmer, C., 
Geller M. 1998, ApJ, 503, 617

\item{\hspace{-0.75cm}} Mateo, M.L. 1998, ARA\&A, 36, 435

\item{\hspace{-0.75cm}} McGaugh, S.S., 1992, PhD Thesis, Univ. Michigan

\item{\hspace{-0.75cm}} McGaugh, S.S., 1996, MNRAS, 280, 337

\item{\hspace{-0.75cm}} Metcalfe, N. , Fong, R. Shanks T. 1995, MNRAS, 274, 769

\item{\hspace{-0.75cm}} Minchin, R.F. 1999, PASA, 16, 12

\item{\hspace{-0.75cm}} O'Neil, K,. Bothun G.D., 2000, ApJ, 529, 811

\item{\hspace{-0.75cm}} Persic, M., Salucci, P., 1992, MNRAS, 258, 14pp

\item{\hspace{-0.75cm}} Petrosian, V. 1998, ApJ, 507, 1

\item{\hspace{-0.75cm}} Phillipps, S., Disney, M. 1986, MNRAS, 221, 1039

\item{\hspace{-0.75cm}} Phillipps, S., Davies, J., Disney, M. 1990, MNRAS, 242, 235

\item{\hspace{-0.75cm}} Phillipps, S., Driver, S.P., 1995, MNRAS, 274, 832

\item{\hspace{-0.75cm}} Phillipps, S., Driver, S.P., Couch, W.J., Smith, R.M., 1998, ApJL, 498, 119

\item{\hspace{-0.75cm}} Pimbblet, K., et al. 2000, (submitted).

\item{\hspace{-0.75cm}} Ratcliffe, A., Shanks, T., Parker, Q., Fong R. 1998, MNRAS, 293, 197

\item{\hspace{-0.75cm}} Sandage, A., Tammann, G.A. 1981, 'A Revised Shapley-
Ames Catalog of Bright Galaxies.' (Washington DC: Carnegie Institute of 
Washington).

\item{\hspace{-0.75cm}} Schechter, P. 1976, ApJ, 203, 297

\item{\hspace{-0.75cm}} Shanks, T. 1990, Proc. IAU Symp. 139 on 
{\it Extragalactic Background Radiation}, Eds S.C. Bowyer, C. Leinert, 
(Publ:Kluwer), 139, 269

\item{\hspace{-0.75cm}} Sodre L(Jr), Lahav O., 1993, MNRAS, 260, 285

\item{\hspace{-0.75cm}} Smith, R.M., Driver, S.P., Phillipps, S. 1997, MNRAS, 287, 415

\item{\hspace{-0.75cm}} Smoker, J.V., Axon, D.J., Davies, R.D. 1999, A\&A, 341,
725

\item{\hspace{-0.75cm}} Sprayberry D., Impey, C., Irwin, M., 1996, ApJ, 463, 535

\item{\hspace{-0.75cm}} Taylor, K., Cannon, R.D., Parker, Q., 1998, in IAU symp. 179, 
Eds.B.J.,McLean., D.A.Golembek, J.J.E.Haynes \& H.E.Payne, (Publ:Kluwer), p 135

\item{\hspace{-0.75cm}} Trentham, N., 1998, MNRAS, 294, 193

\item{\hspace{-0.75cm}} van den Bergh, S., 1998, 'Galaxy Morphology and 
Classification.' (Publ. CUP)

\item{\hspace{-0.75cm}} Williams R.E., {\it et al.} 1996, AJ, 112, 1335

\item{\hspace{-0.75cm}} Willmer, C.N.A., 1997, AJ, 114, 898

\item{\hspace{-0.75cm}} White, S.D.M. \& Rees, M.J. 1978, MNRAS, 183, 341

\item{\hspace{-0.75cm}} White, S.D.M., Tully, B. \& Davis, M. 1988, ApJ, 333, 
L45.
\item{\hspace{-0.75cm}}  Zucca, E., Zamorani, G., Vettolani, G., Cappi, A., Merighi,
R., Mignoli, M., Stirpe, G.M., Macgillivray, H., Collins, C. 
and Balkowski, C., Cayatte, V., Maurogordato, S., Proust, D. 
and Chincarini, G., Guzzo, L., Maccagni, D., Scaramella,
R., Blanchard, A., Ramella, M. 1997, A\&A, 326, 477

\end{description}

\appendix
\section{Testing the isophotal corrections}
It is possible to model several different types of galaxy and compare 
the isophotal magnitude and the total magnitude as calculated in \S 4 with the 
``true'' magnitude. The models are simple, assuming a face on circular galaxy,
composed of a bulge with a de Vaucouleurs $r^{\frac{1}{4}}$ law (de Vaucouleurs
1948) and a disk with an exponential profile (see Eqn.~\ref{eq:exp}). 

\begin{equation}
\mu_{bulge}=\mu_e+1.40+8.325[(r/r_e)^{1/4}-1]
\end{equation}

\noindent where $r_e$ is the half-light radius of the bulge, $\mu_e$ is the
effective surface brightness of the bulge. Here we define it as the mean 
surface brightness within $r_e$\footnote{Note that the term ``effective 
surface brightness'' is sometimes defined as the surface brightness
at $r_e$, for Ellipticals the correction between these definitions of effective
surface brightness is 1.40.}. However, it is more useful to define galaxies
in terms of their luminosities and bulge-to-disk ratios than their effective
radii or disk scale-lengths.

The magnitude of a galaxy and the bulge-disk ratio can be found in terms of 
the above parameters, by:

\begin{equation}
\begin{split}
&M=-2.5\log_{10}[10^{-0.4B}+10^{-0.4D}] \\
&B/T=\frac{B}{B+D}=\frac{1}{1+1/(\frac{B}{D})} \\
&B=\mu_e+1.40-2.5\log_{10}[7.22\pi\,r_e^2] \\
&D=\mu_o-2.5\log_{10}[\pi\,\alpha^2] \\
\end{split}
\end{equation}

\noindent where $B$ is the magnitude of the bulge and $D$ is the magnitude of 
the disk. $B/T$ is the bulge-to-total ratio.
Given the parameters $M$, $B/T$, $\mu_e$ and $\mu_o$, 
a galaxy's light profile is fully defined.

To calculate the difference between the total and the isophotal magnitude it 
is necessary to find the fraction of light lost below the isophote. Since the 
intrinsic detection
isophote varies with the redshift, this difference will be a function
of redshift. For a variety of redshifts from $z=0.001$ to $z=0.201$, the 
fraction of light under the isophote was calculated, by first converting the 
above magnitudes to apparent magnitudes, the intrinsic surface brightnesses to
apparent surface brightnesses and then calculating the scale-lengths 
as above. The conversions from absolute to apparent properties are given in 
Eqn.~\ref{eq:absmag} and Eqn.~\ref{eq:absmu}.

Using $\mu_{lim}=24.67$mag arcsec$^{-2}$, the isophotal radii of the
disk and bulge are calculated.

\begin{equation}
\begin{split}
&r_{iso,d}=(\frac{\mu_{lim}-\mu_o}{1.086})\,\alpha \\
&r_{iso,b}=(\frac{\mu_{lim}-(\mu_e+1.40)}{8.325}+1)^4\,r_e \\
\end{split}
\end{equation}

The fraction of light above the isophote is then calculated using the equation
below.

\begin{equation}
f=1-\sum_{n=0}^{2\beta-1}\frac{1}{n!}g^n\,e^{-g}
\end{equation}

\noindent where $\beta$ is the de Vaucouleur's parameter, which is 1 for a disk
and 4 for a bulge. $g=7.67(\frac{r}{r_e})^{1/4}$ in bulges and 
$g=\frac{r}{\alpha}$ in disks. The isophotal magnitude and isophotal radius 
of the galaxy can now be calculated.

\begin{equation}
\begin{split}
&m_{iso,b}=m_b-2.5\log_{10}\,f_b \\
&m_{iso,d}=m_d-2.5\log_{10}\,f_d \\
&m_{iso}=-2.5\log_{10}[10^{-0.4m_{iso,b}}+10^{-0.4m_{iso,d}}] \\
&r_{iso}=\rm{max}(r_{iso,b},r_{iso,d}) \\
\end{split}
\end{equation}

Now that the isophotal magnitude, $m_{ISO}$, has been found it is possible to 
convert it back to an absolute magnitude, $M_{ISO}$.
The isophotal magnitude and radius 
are fed now back through the equations in \S 4 and a value of $m^{corr}$ is 
calculated. This is converted to an absolute magnitude and 
Table 4 shows a comparison of $M_{true}$, 
$M_{iso}$ and $M_{corr}$ at $z=0.12$ for the main Hubble Type 
galaxies and low surface brightness galaxies. 
The properties of the LSBGs ($B/T$, $\mu_e$ and $\mu_o$) come from averaging 
the B-band data in Beijersbergen, de Blok \& van der Hulst (1999), the 
values of $B/T$ for S0, Sa, Sb and Sc galaxies are taken from 
Kent (1985). The $\mu_e$ values 
are taken from Fig. 5 of Kent (1985) for Ellipticals and Spirals, 
extrapolating where necessary (i.e. for Sb, take a value between that median 
for Sa-Sb and Sbc+) and subtracting 1.40 for the conversion from the surface 
brightness at $r_e$ to the effective surface brightness. For $\mu_o$, we used
the Freeman value for Spiral disks (Freeman, 1970). Irregular galaxies are 
usually placed beyond Spirals on the Hubble Sequence. They have either no bulge
or a small bulge (van den Bergh, 1998; Smoker, Axon \& Davies, 1998) and the 
majority are small $M_B>-17$. Ferguson \& Binggeli (1994) show that all 
galaxies with $M_B>-16$ can be fitted with exponential profiles. 
  
The absolute magnitude is kept constant at $M_{true}=-21.$ and the values of
$M_{iso}$ and $M_{corr}$ are calculated at $z=0.12$. Adjusting the value of
$M_{true}$ does not affect the changes in magnitude at any particular redshift,
but does effect the maximum redshift that the galaxy can be seen at. The 
differences between $M_{iso}$ \& $M_{true}$ and $M_{corr}$ \& $M_{true}$ both
increase with redshift. In all cases apart from the Elliptical galaxy, the
value of $M_{corr}$ is a {\it substantial} improvement over $M_{iso}$ and 
remain below the photometric error for all types out to $z=0.12$.

\section{Visibility Theory}

The equations below are reproduced from Phillipps, Davies \& Disney (1990) and
Disney \& Phillipps (1983). These are used to calculate the volume over which a
galaxy of absolute magnitude $M$, and central surface brightness $\mu_o$ can be
seen. The theory determines the maximum distance that a galaxy can be seen to,
using two constraints: the apparent magnitude that the galaxy would have, and 
the apparent size that the galaxy would have. The first constraint sets a 
limit on the luminosity distance to the galaxy which is the distance that a 
galaxy is at when it becomes too faint to be seen. 

\begin{equation}
d_1=[f(\mu_{lim}-\mu_o)]^{\frac{1}{2}}\,10^{[0.2(m_{lim}-M-25-K(z))]}\,\rm{Mpc}
\end{equation}
 
where $f(\mu_{lim}-\mu_o)$ is the fraction of light above the isophotal 
detection threshold and is profile dependent. For a spiral disk with an 
exponential profile, the fraction of light is:

\begin{equation}
f(\mu_{lim}-\mu_o)=1-[1+0.4\ln(10)(\mu_{lim}-\mu_o)]\,
10^{[-0.4(\mu_{lim}-\mu_o)]}
\end{equation}

The values of $\mu_{lim}$ and $\mu_o$ have to both be absolute or both be 
apparent, for the relation above to be true. As the known value of $\mu_{lim}$
is the apparent value and the known value of $\mu_o$ is the absolute value, a
redshift dependent factor must be included in all calculations.

\begin{equation}
\mu_{lim}-\mu_o \rightarrow \mu_{lim}^{app}-\mu_o-10\,\log_{10}(1+z)-K(z)
\end{equation}

Thus the maximum distance has a redshift dependence. The luminosity distance is
a function of redshift:

\begin{equation}
d_L(z)=\frac{2c}{H_o}\,[(1+z)-(1+z)^{+0.5}]
\end{equation}

The maximum distance can be found numerically by, for instance, a Newton- 
Raphson iteration as 

\begin{equation}
d_1(z)-d_L(z) = 0 \label{eq:NR1}
\end{equation}

\noindent at the maximum distance. 

The second constraint, the size limit is found by a similar methodology. The
formulation of the size limit is:

\begin{equation}
d_2=C\,g(\mu_{lim}-\mu_o)\,10^{[0.2(\mu_o-M)]}\,/\theta_{lim}\,\rm{Mpc}
\end{equation}

where C is a profile dependent constant. $g(\mu_{lim}-\mu_o)$ is the isophotal
radius in scale lengths. $\theta_{lim}$ is the minimum apparent diameter.
For a spiral disk with an exponential profile:

\begin{equation}
C=\frac{2}{\sqrt{2\pi}}\,10^{-5}
\end{equation}
\begin{equation}
g(\mu_{lim}-\mu_o)=0.4\ln(10)[\mu_{lim}^{app}-\mu_o-10\,\log_{10}(1+z)-K(z)]
\end{equation}

In this case $d_2$ is an angular-diameter distance, not a luminosity distance.

\begin{equation}
d_A(z)=\frac{2c}{H_o}\,[(1+z)^{-1}-(1+z)^{-1.5}]
\end{equation}

\begin{equation}
d_2(z)-d_A(z) = 0 \label{eq:NR2}
\end{equation}

Once the redshifts $z_1$ and $z_2$, which are the solutions of 
Eqn~\ref{eq:NR1} and Eqn~\ref{eq:NR2} , have been found, the 
maximum redshift which the galaxy can be seen to is the minimum of $z_1$ and 
$z_2$.

\begin{equation}
z_{max} = \rm {min} (z_1,z_2)
\end{equation}

For the 2dFGRS, the parameters above are: $\mu_{lim}=24.67$mag arcsec$^{-2}$, 
$m_{lim}=19.45$mag, and $\theta_{lim}=7.2''$. $z_{max} \leq 0.12$, is another 
limit imposed on this survey. 

There are also limits on the minimum redshift. These are also caused by 
selection effects in the survey. One limit comes from the maximum apparent 
brightness of galaxies in the sample, which is $m=14.00$. The minimum 
redshift, $z_3$ is calculated just as $z_1$. There is a also a maximum radius
for galaxies, which is $\theta_{max}=200''$. This comes from the sky 
subtraction process. This leads to $z_4$, calculated in the same way as $z_2$.

\begin{equation}
z_{min} = \rm {max} (z_3,z_4)
\end{equation}

$z_{min} \geq 0.015$. This limit is to prevent peculiar 
velocities dominating over the Hubble flow velocity.

\begin{equation}
V(M,\mu_o)=\int_{z_{min}(M,\mu_o)}^{z_{max}(M,\mu_o)}\,\frac{\sigma\,c
d_L^2}{H_o\,(1+z)^{3.5}}dz
\end{equation}

\noindent where $\sigma$ is the area on the sky in steradians, c is the speed
of light and $H_o$ is Hubble's constant.
The visibility $V(M,\mu_o)$ represents the volume over which a spiral disk 
galaxy of absolute magnitude $M$ and central surface brightness $\mu_o$ can 
be observed. The central surface brightness can be converted to effective
surface brightness using the following formula.

\begin{equation}
\mu_e = \mu_o + 1.124
\end{equation}

\section{Tables*}

\begin{table*}
\caption{A comparison of the local luminosity density from
recent magnitude-limited redshift surveys.}
\begin{tabular}{|l|l|c|c|c|c} \hline$^{*2}$
Survey      & Reference & $M^*$ & $\phi^*$ & $\alpha$ & $j_{B}/10^{8}h_{100}
L_{\odot}$Mpc$^{-3}$ \\	\hline
\hline
SSRS2       & Marzke {\it et al.} (1998) & -19.43 & 1.28$\times$10$^{-2}$ & -1.12 & 1.28  		  \\  	
Durham/UKST & Ratcliffe {\it et al.} (1998) & -19.68 & 1.7$\times$10$^{-2}$ & -1.04 & 2.02  	\\  
ESP 	    & Zucca {\it et al.} (1997) & -19.61 & 2.0$\times$10$^{-2}$ & -1.22 & 2.58	 \\  
LCRS$^{*}$  & Lin {\it et al.} (1996) & -19.19 & 1.9$\times$10$^{-2}$ & -0.70 & 1.26	\\  
EEP         & Efstathiou {\it et al.} (1988) & -19.68 & 1.56$\times$10$^{-2}$ & -1.07 & 1.89	  \\  
Stromlo/APM & Loveday {\it et al.} (1995) & -19.50 & 1.40$\times$10$^{-2}$ & -0.97 & 1.35	  \\  
Autofib	    & Ellis {\it et al.} (1996) & -19.20 & 2.6$\times$10$^{-2}$ & -1.09 & 2.05  \\  
CfA$^{*2}$  & Marzke {\it et al.} (1994) & -19.15 & 2.4$\times$10$^{-2}$ & -1.00 & 1.71	  \\	
\end{tabular}

\noindent
$^{*}$ The LCRS used a Gunn r filter. The value of $M^*$ has been converted to
$M_{B}$ using $<b_j-R>_o = 1.1$ for the Johnson B-band (Lin et al. 1996).
$^{*2}$ The CfA used Zwicky Magnitudes. The value of $M^*$ has been 
converted to $M_{B}$ using $<b_j-M_Z>_o = -0.35$ for the Johnson B-band and 
$\phi^*$ has been reduced by 60\% (Gaztanaga \& Dalton (2000).
\end{table*}

\begin{table*}
\caption{A table of the number density as a function of $M$ and $\mu_e$ for 
the 2dFGRS data. The values are in units of $1.0\times10^{-4}$galaxies 
Mpc$^{-3}$. The dashes are in bins where visibility theory says that the 
volume sampled, $V(M,\mu)<10^4$Mpc$^3$. The bold numbers are in bins where 
there are more than 25 galaxies. The italic numbers are in bins where there
are less than 25 galaxies and visibility theory is used to calculate the 
number density. Where this becomes less than 1 galaxy in $10^4$Mpc$^3$, the
limit $<1$ is used.}
\begin{tabular}{l||cccccccccc} \hline
Bin & 20.1 & 20.6 & 21.1 & 21.6 & 22.1 & 22.6 & 23.1 & 23.6 & 24.1 & 24.6 \\ 
\hline \hline
-24.0 & --- & --- & --- & $\it {<1}$ & $\it {<1}$ & $\it {<1}$ & $\it {<1}$ & $\it {<1}$ & $\it {<1}$ & $\it {<1}$ \\
-23.5 & $\it {<1}$ & $\it {<1}$ & $\it {<1}$ & $\it {<1}$ & $\it {<1}$ & $\it {<1}$ & $\it {<1}$ & $\it {<1}$ & $\it {<1}$ & $\it {<1}$ \\
-23.0 & $\it {<1}$ & $\it {<1}$ & $\it {<1}$ & $\it {<1}$ & $\it {<1}$ & $\it {<1}$ & $\it {<1}$ & $\it {<1}$ & $\it {<1}$ & $\it {<1}$ \\
-22.5 & $\it {<1}$ & $\it {<1}$ & $\it {<1}$ & $\it {<1}$ & $\it {<1}$ & $\it {<1}$ & $\it {<1}$ & $\it {<1}$ & $\it {<1}$ & $\it {<1}$ \\
-22.0 & $\it {<1}$ & $\it {<1}$ & $\it {<1}$ & $\it {<1}$ & $\it {<1}$ & $\it {<1}$ & $\it {<1}$ & $\it {<1}$ & $\it {<1}$ & $\it {<1}$ \\
-21.5 & $\it {<1}$ & $\it {<1}$ & $\it {<1}$ &{\bf 1.4$\pm$0.4} & $\it {<1}$ & $\it {<1}$ & $\it {<1}$ & $\it {<1}$ & $\it {<1}$ & $\it {<1}$ \\
-21.0 & $\it {<1}$ & $\it {<1}$ &{\bf 4.2$\pm$0.6} &{\bf 6.1$\pm$0.8} &{\bf 3.3$\pm$0.6} & $\it {<1}$ & $\it {<1}$ & $\it {<1}$ & $\it {<1}$ & $\it {<1}$ \\
-20.5 & $\it {<1}$ &{\bf 1.9$\pm$0.4} &{\bf 8.8$\pm$0.9} &{\bf 18$\pm$1.3} &{\bf 12$\pm$1.1} &{\bf 3.0$\pm$0.5} & $\it {<1}$ & $\it {<1}$ & $\it {<1}$ & $\it {<1}$ \\
-20.0 & $\it {<1}$ &{\bf 2.9$\pm$0.7} &{\bf 14$\pm$1.2} &{\bf 42$\pm$2.0} &{\bf 33$\pm$1.8} &{\bf 9.0$\pm$0.9} & $\it {<1}$ & $\it {<1}$ & $\it {<1}$ & $\it {<1}$ \\
-19.5 & $\it {<1}$ & $\it {<1}$ &{\bf 15$\pm$1.2} &{\bf 59$\pm$2.4} &{\bf 65$\pm$2.5} &{\bf 26$\pm$1.6} &{\bf 5.7$\pm$0.8} & $\it {1.3}$ & $\it {<1}$ & $\it {<1}$ \\
-19.0 & $\it {<1}$ & $\it {1.0}$ &{\bf 13$\pm$1.4} &{\bf 58$\pm$2.4} &{\bf 98$\pm$3.1} &{\bf 52$\pm$2.4} &{\bf 14$\pm$1.4} &{\bf 4.2$\pm$1.0} & $\it {<1}$ & $\it {<1}$ \\
-18.5 & $\it {<1}$ & $\it {<1}$ &{\bf 11$\pm$1.7} &{\bf 61$\pm$3.5} &{\bf 111$\pm$4.4} &{\bf 77$\pm$3.7} &{\bf 35$\pm$2.8} &{\bf 11$\pm$1.9} & $\it {3.0}$ & $\it {<1}$ \\
-18.0 & $\it {<1}$ & $\it {<1}$ &{\bf 15$\pm$4.1} &{\bf 43$\pm$3.5} & {\bf 110$\pm$5.8} &{\bf 113$\pm$6.0} &{\bf 55$\pm$4.6} &{\bf 21$\pm$3.4} & $\it {2.7}$ & $\it {<1}$  \\
-17.5 & $\it {<1}$ & $\it {<1}$ & $\it {2.8}$ &{\bf 31$\pm$4.4} &{\bf 96$\pm$7.4} &{\bf 126$\pm$8.6} &{\bf 62$\pm$6.7} &{\bf 35$\pm$6.9} & $\it {7.1}$ & $\it {<1}$ \\
-17.0 & $\it {2.5}$ & $\it {1.6}$ & $\it {2.2}$ &{\bf 39$\pm$8.9} &{\bf 113$\pm$13} &{\bf 136$\pm$14} & {\bf 103$\pm$14} & {\bf 44$\pm$10} & $\it {20}$ & $\it {<1}$ \\
-16.5 & --- & $\it {<1}$ & $\it {2.3}$ & $\it {27}$ &{\bf 101$\pm$16} &{\bf 151$\pm$19} & {\bf 116$\pm$17} & {\bf 66$\pm$14} & $\it {30}$ & --- \\
-16.0 & --- & --- & --- & $\it {19}$ &{\bf 50$\pm$14} &{\bf 171$\pm$27} &{\bf 130$\pm$24} & $\it {72}$ & --- & --- \\
-15.5 & --- & --- & --- & --- & --- &{\bf 126$\pm$34} & $\it {102}$ & --- & --- & --- \\
-15.0 & --- & --- & --- & --- & --- & --- & --- & --- & --- & --- \\
-14.5 & --- & --- & --- & --- & --- & --- & --- & --- & --- & --- \\
\hline
\end{tabular}
\end{table*}

\begin{table*}
\small
\caption{A summary of the luminosity densities and upper limits on $\Omega_M$
found from the 2dFGRS data.}
\begin{tabular}{|l|c|c|c|} \hline
Parameters  & $j/10^{8}hL_{\odot}$Mpc$^{-3}$ & $\Omega_M$ \\	
\hline
\hline
$\mu_{lim}=24.67$  & $2.49\pm0.20$ & $0.24\pm0.05$ \\
$\mu_{lim}=24.97$  & $2.33\pm0.17$ & $0.23\pm0.05$  	\\  
$\mu_{lim}=24.37$  & $3.00\pm0.26$ & $0.29\pm0.06$ 	 \\  
Simple LF & $1.82\pm0.09$ & $0.18\pm0.04$  \\ 
\hline
\end{tabular}
\end{table*}

\begin{table*}
\small
\caption{A summary of the errors in the isophotal and corrected magnitudes.}
\begin{tabular}{|l|c|c|c|c|c|c|} \hline
Hubble Type  & $M_{true}$ & $B/T$ & $\mu_e$ & $\mu_o$ & $M_{iso}$ & $M_{corr}$ \\	
\hline
\hline
E & -21.00 & 1.00 & 20.5 & --- & -21.00 & -21.07 \\
S0 & -21.00 & 0.65 & 19.2 & 21.7 & -20.83 & -20.89 \\
Sa & -21.00 & 0.50 & 19.6 & 21.7 & -20.75 & -20.86 \\
Sb & -21.00 & 0.30 & 20.4 & 21.7 & -20.64 & -20.86 \\
Sc & -21.00 & 0.15 & 21.0 & 21.7 & -20.54 & -20.89 \\
Sd & -21.00 & 0.10 & 21.2 & 21.7 & -20.51 & -20.92 \\
LSBG & -21.00 & 0.12 & 27.7 & 23.0 & -19.08 & -20.84 \\
Irr & -21.00 & 0.00 & --- & 22.7 & -19.67 & -21.00 \\
\hline
\end{tabular}
\end{table*}

\end{document}